\documentclass[a4paper,fleqn,usenatbib]{mnras}
\usepackage[T1]{fontenc}
\usepackage{ae,aecompl}

\usepackage{graphicx}	
\usepackage{amsmath}	
\usepackage{amssymb}	

\def\apjs{{\rm ApJS}}                  
\def\apjl{{\rm ApJ Let}}               
\def\aj{{\rm AJ}}                      
\def\mnras{{\rm MNRAS}}                        

\def\Msol{\thinspace\hbox{$\hbox{M}_{\odot}$}}

\def\a4{\hsize 17.0cm \vsize 25.cm}
\newcommand{\der}[2]  { \frac{{\rm d}#1}{{\rm d}#2} }

\title[Gas expulsion vs gas retention in YSCs]{Gas expulsion vs gas retention 
in young stellar clusters II: effects of cooling and mass segregation}

\author[S. Silich and G. Tenorio-Tagle]
{Sergiy Silich \thanks{E-mail: silich@inaoep.mx}
and Guillermo Tenorio-Tagle
\\
Instituto Nacional de Astrof\'\i sica \'Optica y Electr\'onica, AP 51, 
      72000 Puebla, M\'exico\\
}

\date{Accepted XXX. Received YYY; in original form ZZZ}

\pubyear{2018}

\begin{document}
\label{firstpage}
\pagerange{\pageref{firstpage}--\pageref{lastpage}}
\maketitle

\begin{abstract}
Gas expulsion or gas retention is a central issue in most of the models for
multiple stellar populations and light element anti-correlations in globular
clusters.  The success of the residual matter expulsion or its retention
within young stellar clusters has also a fundamental importance in order to 
understand how star formation proceeds in present-day and ancient 
star-forming galaxies and if proto-globular clusters with multiple stellar
populations are formed in the present epoch. It is usually suggested  that 
either the residual gas is rapidly ejected from star-forming clouds by 
stellar winds and supernova explosions, or that the enrichment of the residual
gas and the formation of the second stellar generation occur so rapidly, that
the negative stellar feedback is not significant. Here we continue our 
study of the early development of star clusters in the extreme environments 
and discuss the restrictions that strong radiative cooling and 
stellar mass segregation provide on the gas expulsion from dense star-forming 
clouds. A large range of physical initial conditions in star-forming clouds 
which include the star-forming cloud mass, compactness, gas metallicity, star 
formation efficiency and effects of massive stars segregation are discussed. 
It is shown that in sufficiently massive and compact clusters hot shocked 
winds around individual massive stars may cool before merging with their 
neighbors. This dramatically reduces the negative stellar feedback, prevents 
the development of the global star cluster wind and expulsion of the residual 
and the processed matter into the ambient interstellar medium. The critical 
lines which separate the gas expulsion and the gas retention regimes are
obtained.

\end{abstract}


\section{Introduction}
\label{sec1}

Globular clusters (GCs), considered for a long time to be chemically 
homogeneous stellar systems are now confirmed to have distinct stellar
subpopulations with different He contents, different abundances of light
elements and well documented anti-correlations between elements such as
O and Na, Mg and Al \citep[see][and references therein]{Bedin2004,
Gratton2004,Marino2008,Carretta2009,Piotto2012,CabreraZiri2015,
Renzini2015}. It has been proposed that the first generation of stars (1G) is 
formed from the pristine (unpolluted) matter whereas the second (2G) and in 
some cases subsequent stellar generations originate from the leftover gas or 
from accreted primordial matter enriched with the high temperature hydrogen 
burning products produced by the 1G stars 
\citep[see][and references therein]{Gratton2004,Prantzos2006,Decressin2007,
DErcole2008,DeMink2009,Renzini2015,Elmegreen2017}. 
A strong constraint on the multiple populations models is that most GCs
with the exception of the most massive ones are homogeneous in iron-peak
elements \citep[see][and references therein]{Renzini2013}. This implies that
pollution of the pristine gas is not caused by the 1G SN explosions and this 
leads to two major pollution scenarios - fast, and slow.

In the slow pollution scenario stellar winds formed by massive stars and SN 
explosions expel the residual gas from the cluster. Intermediate mass AGB 
stars then shed processed gas at low velocity.  The enriched matter is 
accumulated in the gravitational well of the cluster and mixes with primordial
gas accreted later from the ambient medium. The 2G of stars then results
from this polluted matter \citep[e.g.][]{DErcole2008,DErcole2010,Conroy2011}.  

In the fast pollution scenario it is suggested that the 2G stars are formed
from the matter polluted within a short time-scale either by fast rotating 
massive stars \citep{Prantzos2006,Decressin2007} or by interacting massive 
binaries \citep{DeMink2009}. In different versions of this scenario it has 
been suggested that proto-stellar discs sweep up matter enriched by fast 
rotating massive stars or interacting massive binaries \citep[Early Disc 
Accretion Model][]{Bastian2013,Cassisi2014}. Other authors have proposed 
that fast rotating massive stars form slow winds and trigger the 2G formation 
in their vicinity \citep{Decressin2007}, or that 
pollution occurs due to interactions between massive stars in very dense 
stellar clusters \citep{Elmegreen2017}. In the rapidly cooling star cluster 
wind model by \citet{Wunsch2017} dense enriched clumps are formed in the 
central zones of thermally unstable star cluster winds. They then 
self-shield against the ionizing photons and accumulate inside the cluster to 
form the 2G of stars before massive stars begin to explode as SNe.  

Gas expulsion or gas retention is a central issue in all of these models. It 
is also crucial in order to understand how stellar feedback affects the 
residual gas in young present-day clusters and in those formed in assembling
galaxies and if proto-globular clusters with multiple stellar populations
can be formed at the present epoch.

There is a common belief that negative feedback provided by massive stars 
rapidly clears out star-forming regions from the residual gas and that this 
process acquires even larger importance in denser, more compact clusters with 
smaller mean separations between neighboring stars \citep[e.g.][]{Krause2012}.
In most multiple populations models the negative stellar feedback either is 
not discussed, or it is assumed that it results in a star cluster wind. The 
full 3D numerical simulations by \citet[][]{Calura2015} seem to confirm these 
expectations. 

Recent attempts to find the residual gas in present-day young massive clusters
did not reveal any significant amount of gas (see Bastian et al. 2013, 
2014). However observations in radio, IR and recently in the sub-millimeter 
regime have revealed several very bright, optically obscured objects which are 
believed to be massive, very compact star clusters still embedded into their 
natal star-forming clouds \citep[][]{Turner2000,Gorjian2001,Turner2003,
Beck2012,Whitmore2014,Beck2015,Turner2015,Consiglio2016,Turner2017,Oey2017}. 
It is not clear so far if in these cases we witness the earliest stages of 
massive clusters assembling as it is often suggested in the case of the 
still enshrouded clusters, or if the negative star formation feedback is in
these cases strongly suppressed by the high-pressure environment or by
in-falling gas filaments \citep[see the discussion in][]
{MartinHernandez2005,Silich2007,Whitmore2014,Matzner2015,Calzetti2015,
Smith2016,Silich2017,Consiglio2017}.

This led us to discuss the impact that turbulent pressure and strong radiative 
cooling provide on the gas expulsion from clusters formed under extreme 
conditions.
In \citet[][paper I]{Silich2017} we have shown that in very massive and
compact clusters wind-driven shells around individual massive stars do not
merge, suppressing the development of a global star cluster wind. Here
we discuss the post-stalling hot shocked gas evolution. It is shown that
hot shocked gas zones around individual massive stars then grow in the 
subsonic regime until they merge with hot bubbles produced by neighboring
massive stars or until catastrophic cooling sets in and the shocked gas
becomes thermally unstable. A global star cluster wind that expels the 
residual gas from the star-forming region is formed in the former case, 
whereas in the latter case the negative stellar feedback is strongly 
suppressed, hot shocked zones around individual stars do not merge and the 
star cluster wind is not formed. The critical lines which separate stellar 
clusters which expel or retain the residual gas and the matter returned by 
massive stars are derived. We also discuss how the initial mass segregation 
affects the critical lines.

We consider instantaneous star formation and depart from the basic 
assumption that stellar winds are synchronized and continuous while SNe do not
coincide either in space or time. Therefore we do not examine the effects of
SNe explosions assuming that the supernovae products blow away from clusters 
providing little effect on the leftover gas distribution 
\citep{TenorioTagle2015} or that most massive stars do not explode, but 
collapse directly into black holes as has been argued by other authors
\citep[see][and references therein]{Decressin2010,Adams2017A,Adams2017B,
Mirabel2017A,Mirabel2017B}. 
 
The paper is organized as follows: in section 2 we discuss the stellar and
gas density distribution in star-forming molecular clouds, used later
on for our numerical calculations. Here we also present the adopted global
and individual stellar wind energy and mass lost rate. In section 3 the
distributions of pressure, density, temperature and expansion velocity in the 
pressure-confined shocked wind zones around individual stars are obtained by 
solving the set of the steady state spherically-symmetric hydrodynamic 
equations numerically. It is demonstrated that in dense, compact clusters
hot shocked winds may cool down before merging with their neighbors. In 
section 4 the results from these simulations are used to build critical lines
which separate clusters which are able to retain all gas, which includes the 
residual gas and that reinserted by massive stars, from those which form 
global star cluster winds, clear their star-forming clouds and expel all gas 
into the ambient interstellar medium. Our major results are summarized in 
Section 5.    

\section{Model setup}
\label{sec2}

\subsection{The star-forming cloud model}

Following recent papers \citep[e.g.][]{Krause2012,Calura2015,Silich2017} we 
consider star-forming clouds with a Plummer density distribution: 
\begin{eqnarray}
      \label{eq.1a}
      & & \hspace{-1.1cm} 
\rho_g(r) = \frac{3 (1 - \epsilon) M_{tot}}{4 \pi a^3} \left(1 + \frac{r^2}{a^2}\right)^{-5/2} ,
      \\[0.2cm] \label{eq.1b}
      & & \hspace{-1.1cm} 
\rho_{\star}(r) = \frac{3 \epsilon M_{tot} }{4 \pi a^3} \left(1 + \frac{r^2}{a^2}\right)^{-5/2} ,
\end{eqnarray}
where $\rho_g(r)$ and $\rho_{\star}(r)$ are the gas and the stellar mass 
density distribution, respectively, $M_{tot}$ is the initial mass of
the star-forming cloud and $\epsilon$ is the  star formation efficiency. In 
the case of a Plummer density distribution the half-mass radius is $R_{hm} = 
1.3 a$, where $a$ is the characteristic length scale or the core radius of
the star-forming cloud. The pressure profile in such a cloud is determined 
by the equation of the hydrostatic equilibrium \citep{Calura2015}:
\begin{equation}
      \label{eq1b}
\der{P_g}{r} = - \frac{G M(r) \rho_g(r)}{r^2} , 
\end{equation}
where $G$ is the gravitational constant and the total mass $M(r)$ enclosed
within a sphere of radius $r$ is: 
\begin{equation}
      \label{eq1c}
M(r) = \frac{r^3 M_{tot}}{(r^2 + a^2)^{3/2}} . 
\end{equation}
One can integrate equation (\ref{eq1b}) and obtain the distribution of 
the gas pressure in the star-forming cloud:
\begin{equation}
      \label{eq1d}
P_g(r) = \frac{(1 - \epsilon) G M^2_{tot}}{8 \pi a^4} \left(1 + 
         \frac{r^2}{a^2}\right)^{-3} + C  , 
\end{equation}
where $C$ is the integration constant. Hereafter we assume that at large 
distances from the cloud center the gas pressure goes to zero and therefore 
$C = 0$. Given the low temperature of the molecular gas, it is likely
that the intra-cluster gas pressure is dominated by turbulence 
\citep{Elmegreen1997,Elmegreen2017,Johnson2015}. In this case $P_g = 
\rho_g \sigma^2$ \citep[e.g.][]{Smith2006}, where the one-dimensional 
velocity dispersion $\sigma$ is:
\begin{equation}
      \label{eq1e}
\sigma^2 = \frac{G M_{tot}}{6 a} \left(1 + \frac{r^2}{a^2}\right)^{-1/2} .
\end{equation}
The distribution of gas pressure, density and velocity dispersion
$\sigma$ in a $M_{tot} = 10^6$\Msol \, star-forming clouds with a star 
formation efficiency $\epsilon = 0.3$ and different core radii is shown 
in Fig. 1. 
\begin{figure*}
\vspace{16.5cm}
\includegraphics{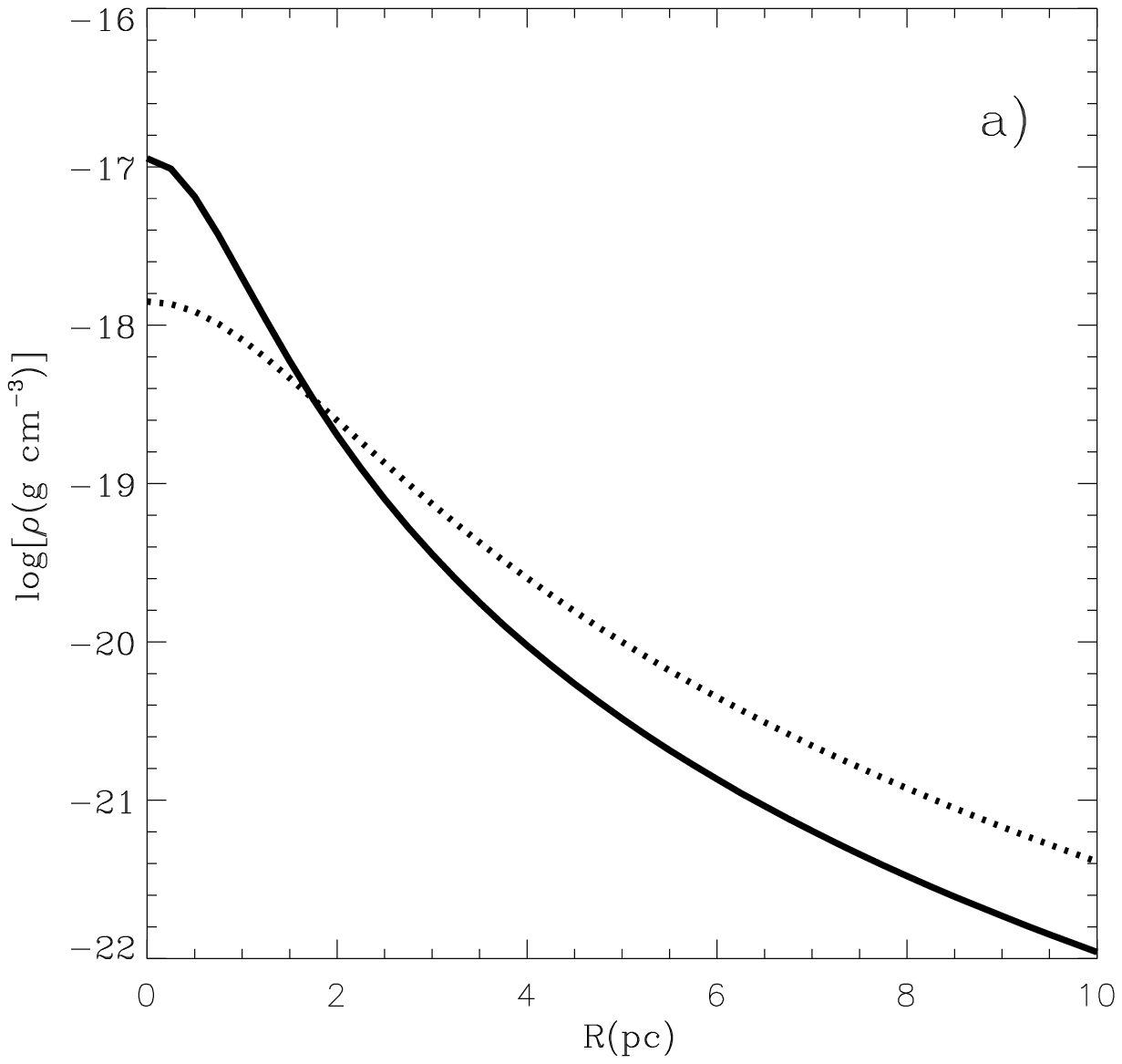}
\includegraphics{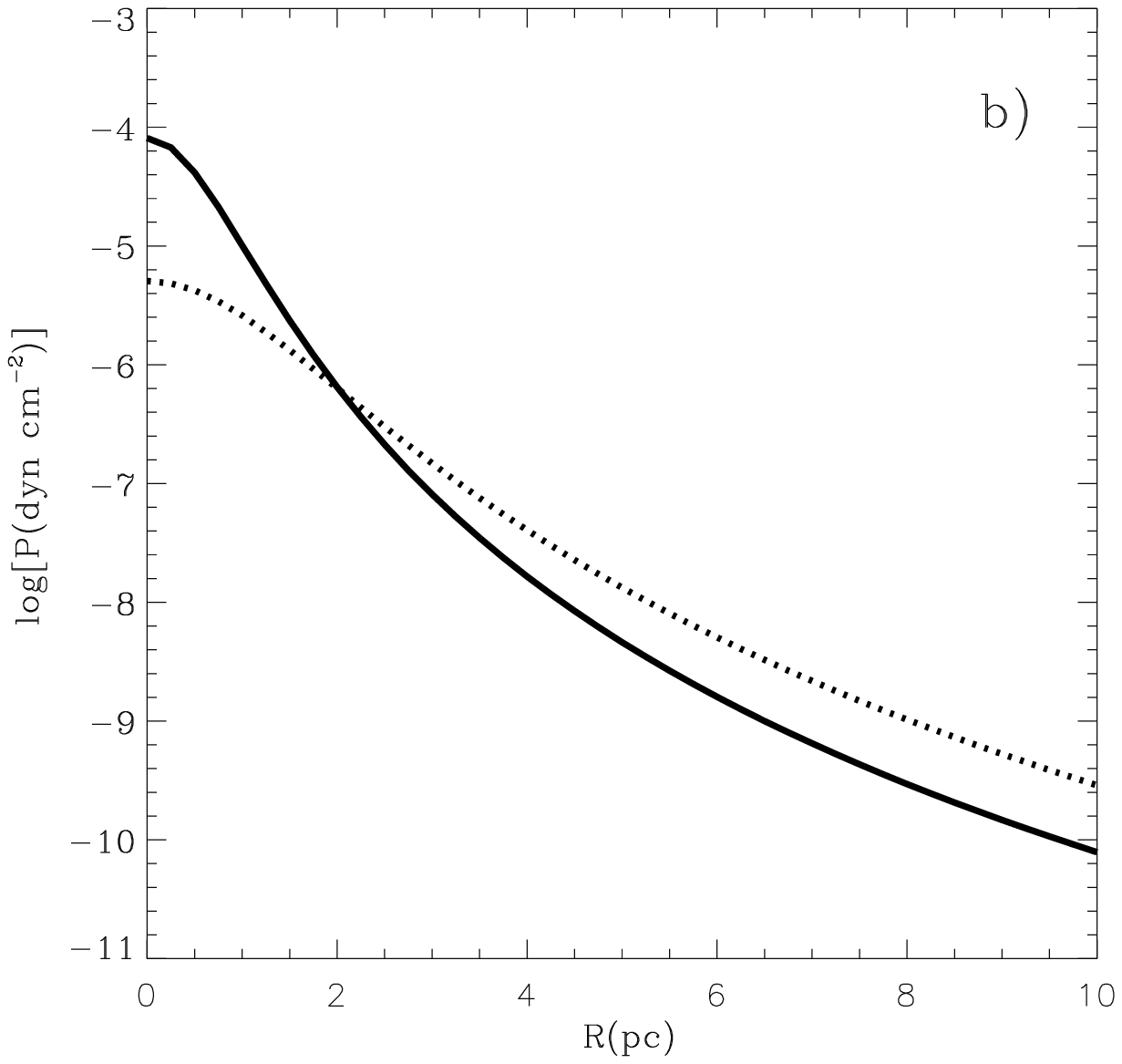}
\includegraphics{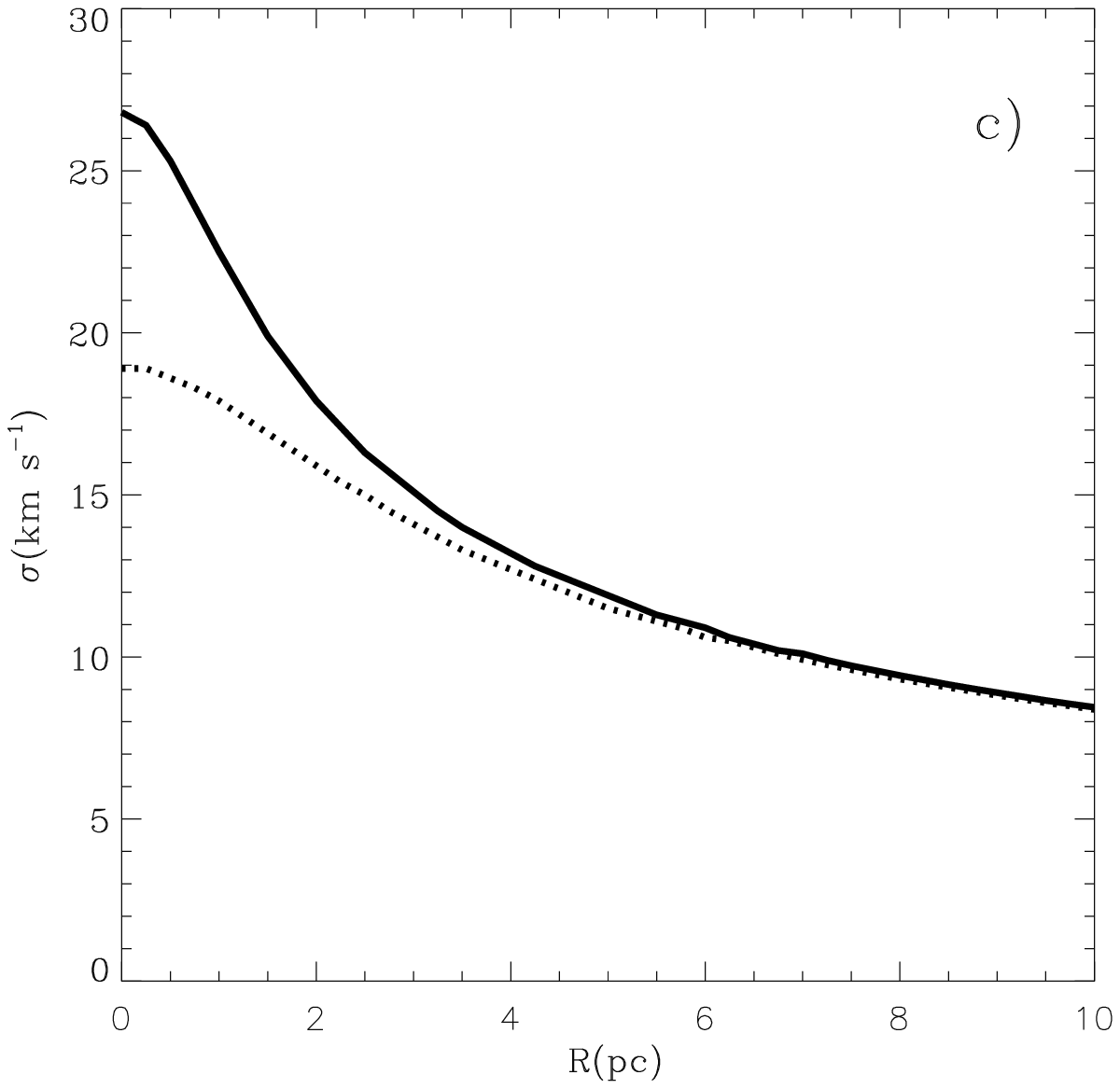}
\caption{The distribution of the gas density (panel a), pressure (panel b) and
         one dimensional velocity velocity dispersion (panel c) in a 
         star-forming cloud. The solid and dotted lines correspond to 
         $10^6$\Msol \, clouds with star formation efficiency $\epsilon = 0.3$
         and core radii $a = 1$pc  and $a = 2$pc, respectively.}
\label{f1}
\end{figure*}
One can note that the central gas density strongly depends on the cloud 
compactness. It is much larger in clouds with smaller core radii (see panel
a in Fig. 1) as in this case the same mass is concentrated in a smaller 
volume. This results in a stronger gravitational pull and a larger gas central
pressure (see panel b in Fig. 1). It is likely that precursors of globular 
clusters were formed in a high pressure environment and that this led to their
compactness and to a large star formation efficiency required to form bound, 
long-lived clusters \citep[see][and references therein]{Ashman2001}. 

Hereafter it is assumed that due to primordial mass segregation all massive 
stars are concentrated within the central zone of the star-forming cloud 
\citep{Baumgardt2008, Dib2008}. The number of massive stars per unit volume 
then is:
\begin{eqnarray}
 \label{eq2a}
      & & \hspace{-1.1cm} 
N(r) = \frac{3 N_{\star}(t)}{4 \pi R_{sg}^3} \left(1 + 
       \frac{R_{sg}^2}{a^2}\right)^{3/2} 
       \left(1 + \frac{r^2}{a^2}\right)^{-5/2} \, r < R_{sg} , 
      \\[0.2cm]     \label{eq2b}
      & & \hspace{-1.1cm}
N(r) = 0 \quad r > R_{sg} ,
\end{eqnarray}
where $N_{\star}(t)$ is the total number of massive stars in the cluster and
$R_{sg}$ is the radius of the mass-segregation zone. In this
case the mean separation between nearby massive stars $\Delta = 2 X$, where
the half-distance between neighboring sources is:
\begin{equation}
      \label{eq2c}
X = R_{sg} \left(1 + \frac{R_{sg}^2}{a^2}\right)^{-1/2} 
       \left(1 + \frac{r^2}{a^2}\right)^{5/6} N_{\star}(t)^{-1/3} .
\end{equation}
When the mass segregation radius $R_{sg} \to \infty$, relation (\ref{eq2c})
reduces to that obtained in \citet{Silich2017}, their equation (4). In the 
case of a standard Kroupa initial mass function (IMF) the initial number of 
single massive ($M > 8$\Msol) stars $N_{\star}(0)$ in an instantaneous stellar
cluster scales with the star cluster mass as \citep[e.g.][]{Calura2015}:
\begin{equation}
      \label{eq2d}
N_{\star}(0) = N_0 (M_{\star}/10^6\Msol) ,
\end{equation}
where $N_0 = 10^4$, $M_{\star} = \epsilon M_{tot}$.  Note, that the number of 
massive stars $N_{\star}(t)$ drops whereas the mean separation between 
neighboring massive stars grows after the onset of supernova explosions. At 
this stage 
\begin{equation}
      \label{eq2e}
N_{\star}(t) = N_{\star}(0) - E_{SN} / E_0 ,
\end{equation} 
where $E_{SN}$ is the total energy of SN explosions calculated by means of 
Starburst99 synthetic model \citep{Leitherer1999} and $E_0 = 10^{51}$~erg 
is the energy of each supernova explosion.

It is assumed that the star formation efficiency $\epsilon$ is the same 
at all radii. The possible growth of the star formation efficiency in the
central zone of the star-forming cloud is not considered, but is expected to 
be qualitatively similar to the influence of mass segregation.

\subsection{The stellar wind energy}

Fig. 2 presents the time evolution of the stellar winds cumulative mechanical 
luminosity, mass loss rate and the number of massive stars in an instantaneous
$10^6$\Msol \, cluster with a standard Kroupa IMF and Padova stellar 
evolutionary tracks with AGB stars predicted by the Starburst99 version 7. 
\begin{figure}
\includegraphics[width=\columnwidth]{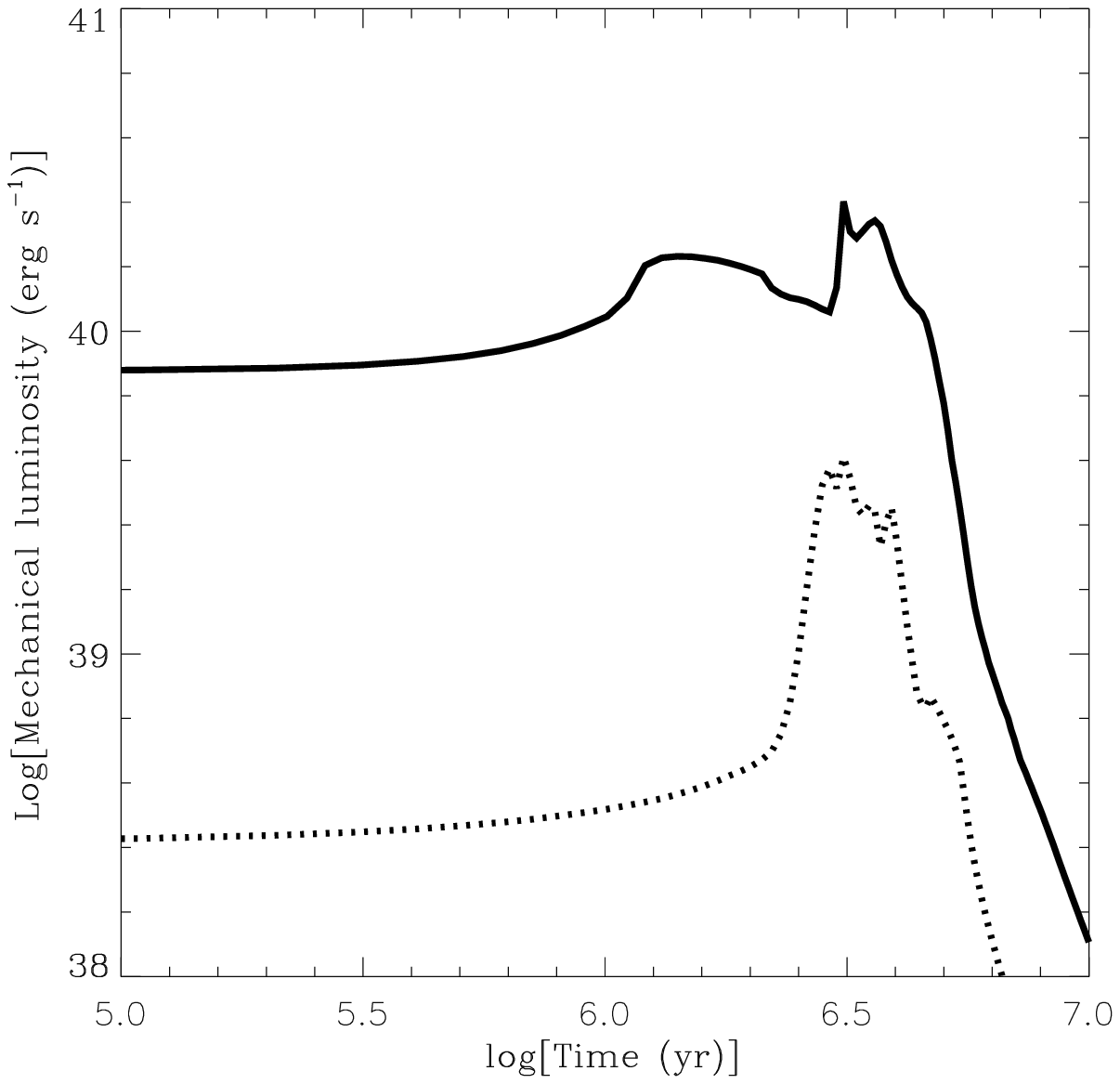}
\includegraphics[width=\columnwidth]{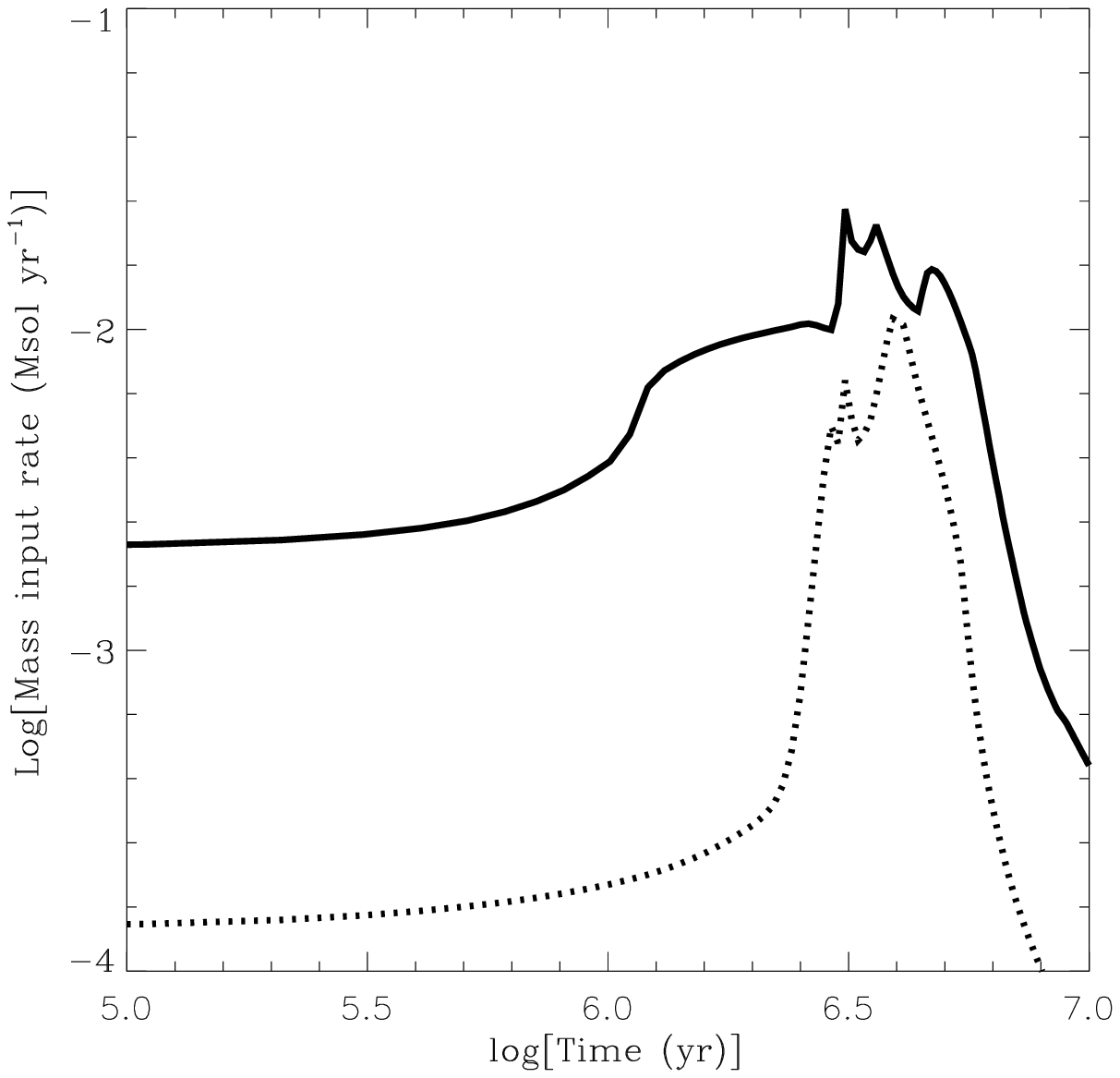}
\includegraphics[width=\columnwidth]{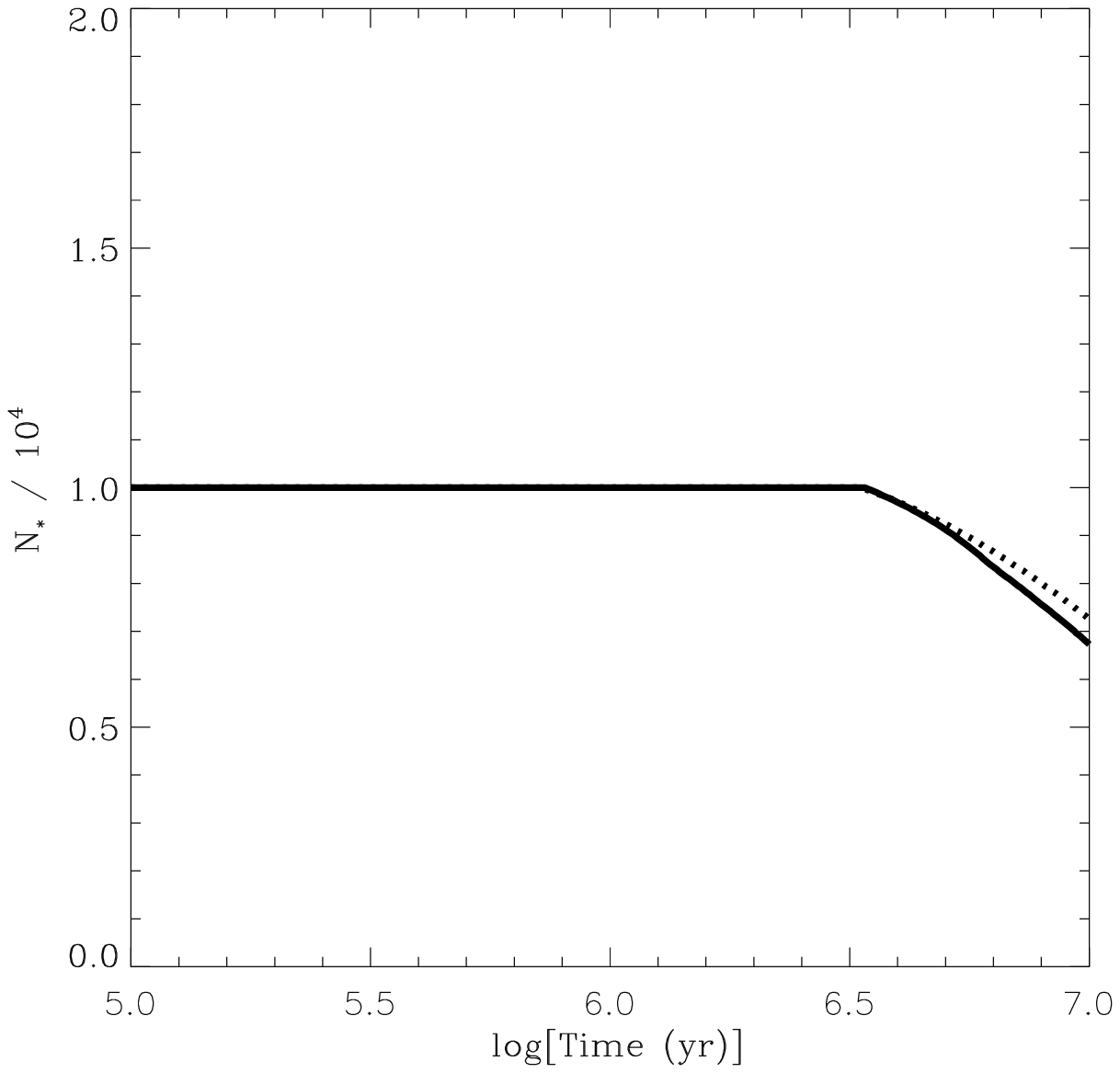}
\caption{Total mechanical luminosity and mass loss rate of stellar winds
         and the number of massive stars in a $10^6$\Msol \, stellar
         cluster. The mechanical luminosity $L_{SC}$ and the stellar mass loss 
         rate ${\dot M}_{SC}$ were calculated by means of the Starburst99, 
         version 7 with Padova stellar evolutionary tracks which includes AGB 
         stars. The number of massive stars was scaled with the star cluster 
         mass and the total energy of SNe. The solid and dotted lines show the
         results of the  calculations for clusters with $Z = Z_{\odot}$ and 
         $Z = 0.02Z_{\odot}$, respectively.} 
\label{f2}
\end{figure}
The solid and dotted lines in Fig. 2 display the mechanical luminosity,
the stellar mass loss rate and the number of massive stars in clusters with 
$Z = Z_{\odot}$ and $Z = 0.02Z_{\odot}$ metallicity, respectively. 
Note that low metallicity 
clusters are much less energetic and return less mass into the ambient medium. 
Hereafter it is assumed that all massive stars are identical. The 
Starburst99 model together with equation (\ref{eq2d}) then allow one to obtain
the mechanical luminosity $L_{\star}$ and the mass loss rate 
${\dot M}_{\star}$ of a typical massive star at any time $t$. During the
first 10~Myr the mechanical luminosities of individual stellar winds 
change within the range $2 \times 10^{34} \le L_{\star} \le 2 \times 
10^{36}$erg s$^{-1}$ in clusters with solar abundances and $3 \times 10^{33} 
\le L_{\star} \le 4 \times 10^{35}$erg s$^{-1}$ in clusters with $Z = 
0.02Z_{\odot}$, respectively, to become negligible after 10~Myr.

\section{Shocked wind bubbles in a high pressure intra-cloud 
            medium}
\label{sec3}

Each massive star produces a fast stellar wind which is heated up at a reverse
shock and drives a leading shock into the ambient medium. The leading shock 
sweeps up the ambient gas into a dense narrow shell which, at first, expands 
supersonically. In paper I we found the critical densities required for 
the wind-driven shells to stall before merging. This occurs when the pressure 
in the hot, shocked wind drops to that in the turbulent ambient medium. The 
wind-driven shell expansion velocity then drops below the intra-cloud gas 
turbulent speed and it begins to disintegrate due to Rayleigh-Taylor 
instabilities and perturbations provided by the ambient gas. However, the hot,
shocked wind zone continuous to grow even after the wind-driven shell stalls 
as the central star continues to deposit mass and energy for a much longer 
time. The hot shocked wind zone grows then in the subsonic regime (see 
Appendix A) until it merges with a neighboring hot shocked wind or reaches its
cooling radius and catastrophic gas cooling sets in. During the evolution the 
stellar wind is heated up at the reverse shock whose radius $R_{RS}$ is 
determined by the condition that the stellar wind ram pressure is equal to the
pressure in the turbulent ambient medium: 
\begin{equation}
      \label{eq4}
R_{RS} = \frac{a^2}{M_{tot}} \left[\frac{4 L_{\star}}
         {(1-\epsilon) G V_{\star}}\right]^{1/2} \left(1 + 
         \frac{R^2}{a^2}\right)^{3/2} ,
\end{equation}
where $V_{\star} = (2 L_{\star}/{\dot M}_{\star})^{1/2}$ is the stellar
wind terminal speed. One can determine how the gas temperature, density and 
pressure are distributed in the slowly growing, subsonic shocked 
wind zones by making use of steady state, spherically symmetric hydrodynamic 
equations which include the gas cooling term \citep[e.g.][]{Silich2004}:
\begin{eqnarray}
      \label{eq5a}
      & & \hspace{-1.1cm} 
\frac{1}{r^2} \der{\rho u r^2}{r} = 0 , 
      \\[0.2cm]   \label{eq5b}  
      & & \hspace{-1.1cm}
\rho u \der{u}{r} = - \der{P}{r} ,
      \\[0.2cm]     \label{eq5c}
      & & \hspace{-1.1cm}
\frac{1}{r^2} \der{}{r}\left[\rho u r^2 \left(\frac{u^2}{2} +
\frac{\gamma}{\gamma-1} \frac{P}{\rho}\right)\right] = -Q ,
\end{eqnarray}
where $Q = n_i n_e \Lambda(T,Z)$ is the cooling rate, $\Lambda(T,Z)$
is the Raymond et al. cooling function \citep{Raymond1976}, $n_i \approx n_e$
are the ion and electron number densities in the hot shocked wind and
Z is the stellar wind metallicity. One can integrate equation (\ref{eq5a})
and present the set of equations (\ref{eq5a})-(\ref{eq5c}) in a form
suitable for the numerical integration:
\begin{eqnarray}
     \label{eq6a}
      & & \hspace{-1.1cm} 
\der{u}{r} = \frac{1}{r \rho} \frac{(\gamma-1) r Q + 2 \gamma u P}
             {u^2 - c^2} ,
      \\[0.2cm]     \label{eq6b}
      & & \hspace{-1.1cm}
\der{P}{r} = -\rho u \der{u}{r} ,
      \\[0.2cm]     \label{eq6c}
      & & \hspace{-1.1cm}
\rho = \frac{{\dot M}_{\star}}{4 \pi u r^2} ,
\end{eqnarray}
where $c^2 = \gamma P / \rho$ is the sound speed in the shocked wind plasma.
The temperature of the shocked wind is $T = \mu_i c^2 / \gamma k$, where
$\mu_i = 14/23 m_H$ is the mean mass per particle in the completely ionized
gas with 1 helium atom per each 10 hydrogen atoms, $m_H$ is the mass of the 
hydrogen atom and $k$ is the Boltzmann constant. The set of equations 
(\ref{eq6a}) - (\ref{eq6c}) should be integrated numerically outwards from 
the reverse shock location. 
\begin{figure*}
\vspace{15.5cm}
\includegraphics{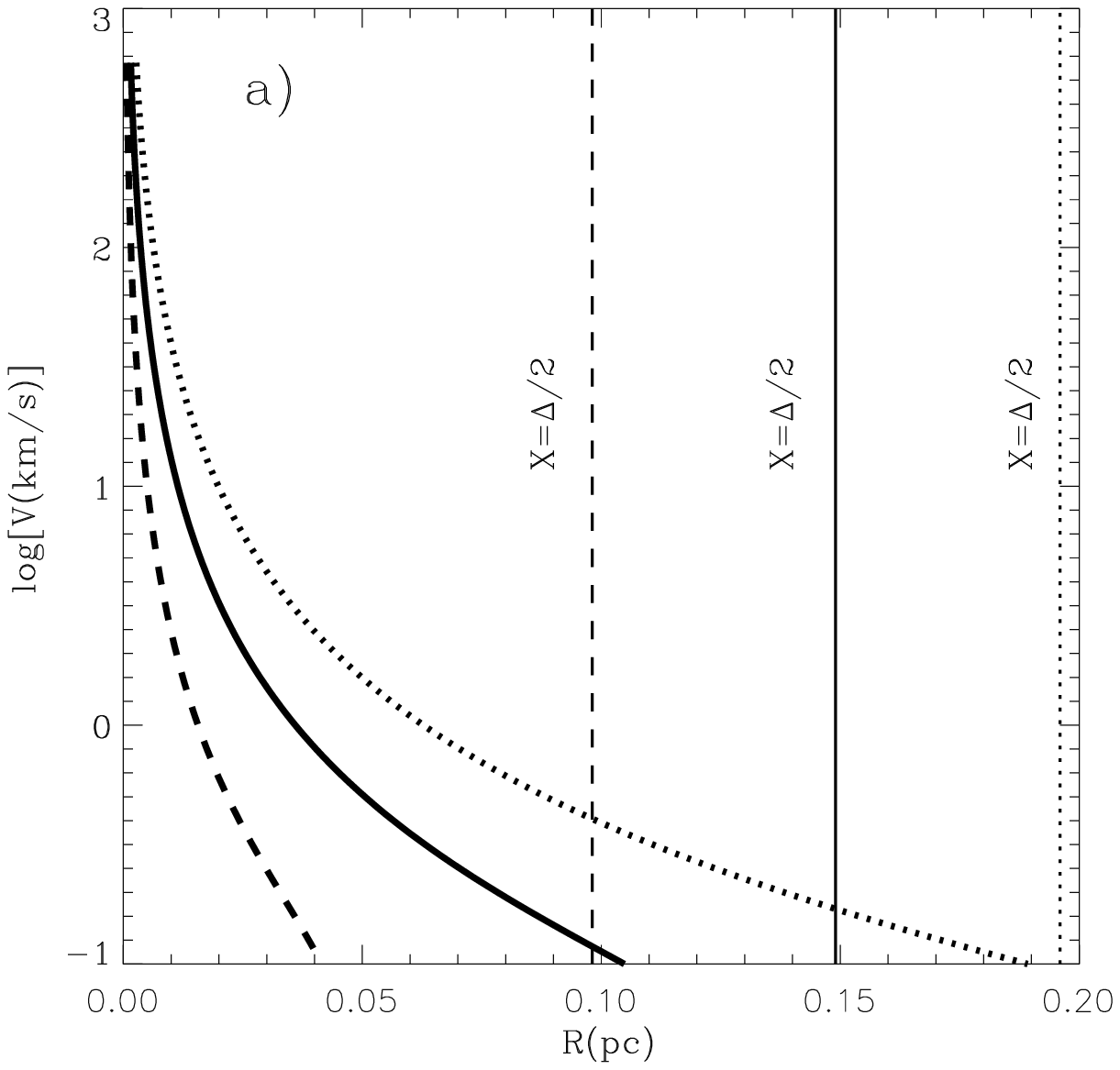}
\includegraphics{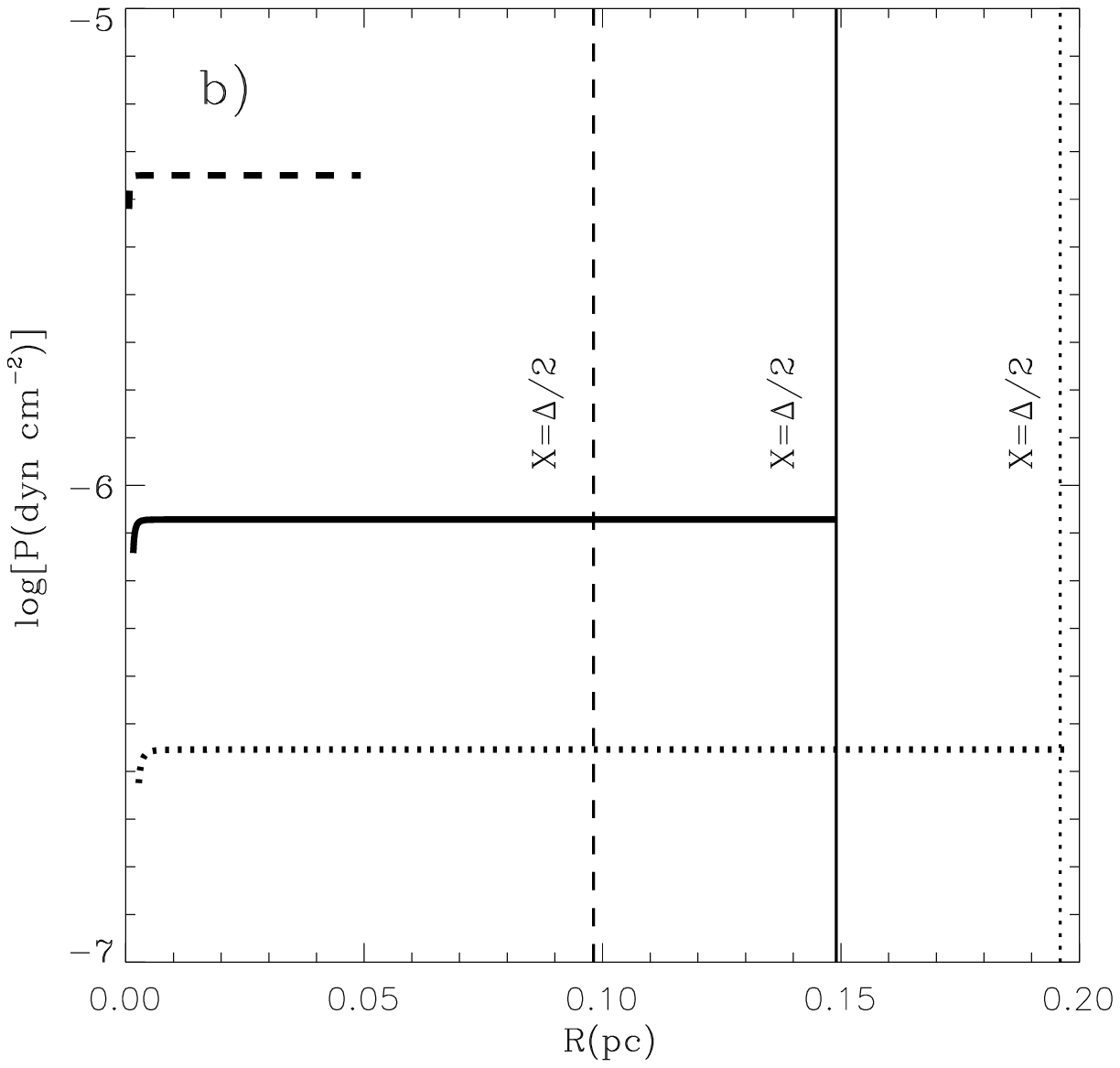}
\includegraphics{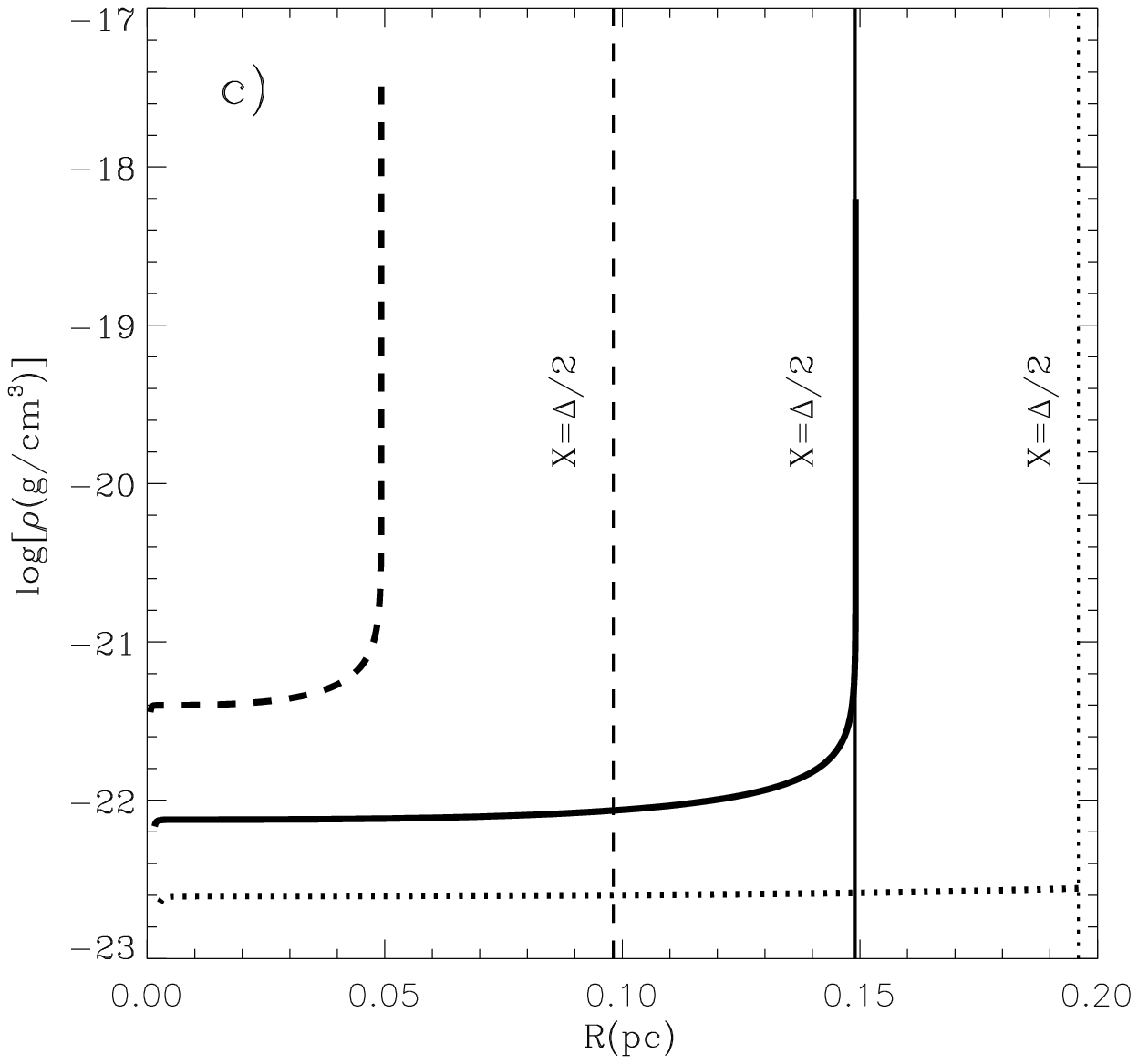}
\includegraphics{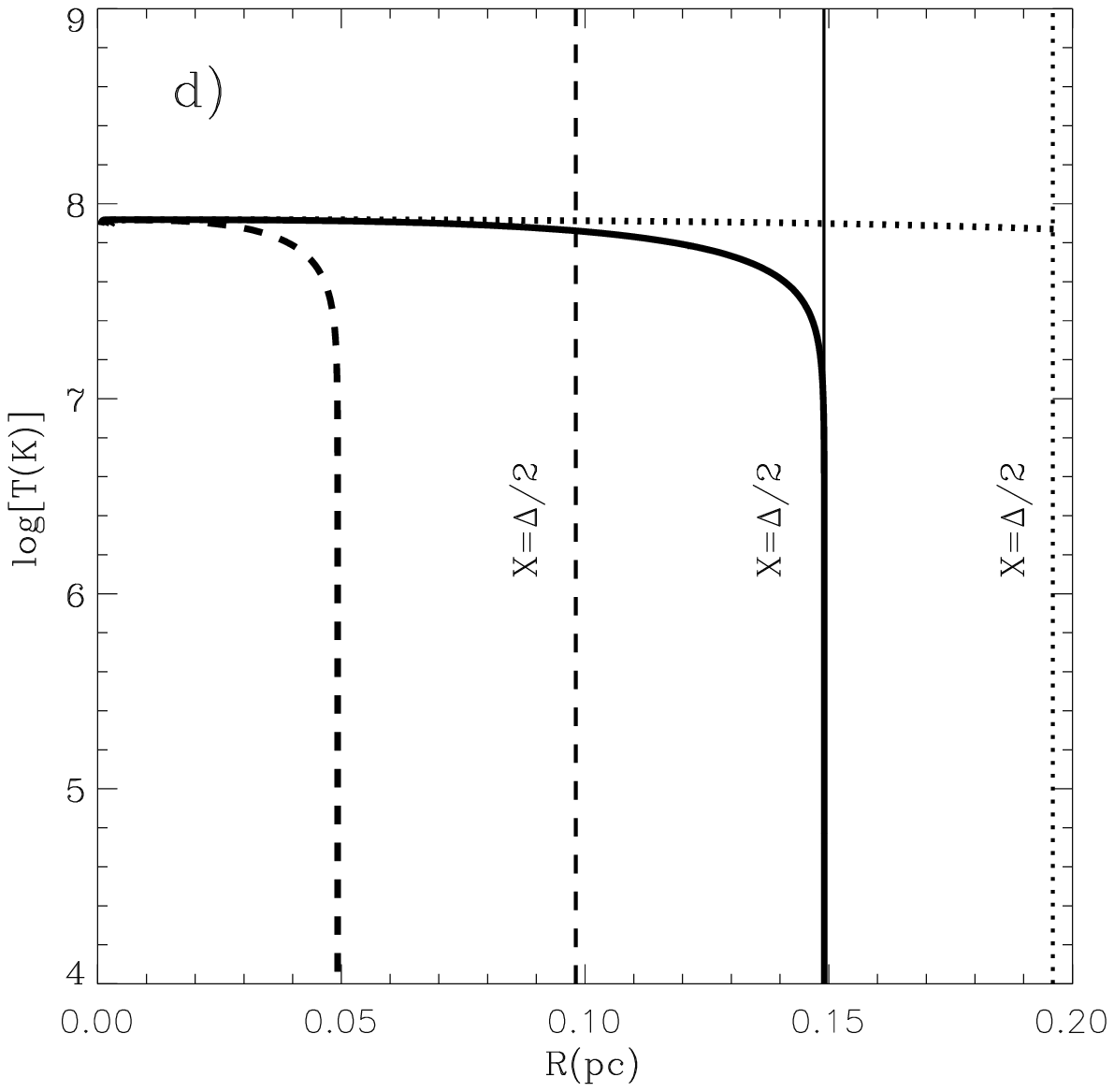}
\caption{The distribution of the gas temperature, velocity, density and
pressure in the shocked wind zone around individual stars located in the
center of a dense star-forming cloud. The calculations were provided for a
a 1Myr old, $3 \times 10^5$\Msol \, cluster formed in a $10^6$\Msol \,
cloud with low metallicity $Z = 0.02Z_{sol}$ and $R_{sg} = a$. The dashed, 
solid and dotted 
lines display the distribution of the expansion velocity (panel a), thermal 
pressure (panel b), density (panel c) and temperature (panel d) in the shocked
wind zone around individual stars in the case when the cluster is formed in 
clouds with different core radii: $a = 2$pc, $a = 3$pc and $a = 4$pc, 
respectively. The vertical dashed, solid and dotted lines show the mean 
half distance between neighboring massive stars in each case.}  
\label{f3}
\end{figure*}
The initial conditions for the numerical integration (the values of the 
shocked wind density, $\rho_{sw}$, temperature, $T_{sw}$, and pressure, 
$P_{sw}$ behind the reverse shock) are determined by the Rankine-Hugoniot 
conditions:
\begin{eqnarray}
      \label{eq7a}
      & & \hspace{-1.1cm} 
\rho_{sw} = \frac{\gamma+1}{\gamma-1} \rho_w(R_{RS}) =
            \frac{\gamma+1}{\gamma-1} \frac{(1-\epsilon) G M^2_{tot}}
            {8 \pi a^4 V^2_{\star}} \left(1 + \frac{r^2}{a^2}\right)^{-3} ,
      \\[0.2cm]     \label{eq7b}
      & & \hspace{-1.1cm}
T_{sw} = \frac{2 (\gamma-1)}{(\gamma+1)^2} \frac{\mu_i}{k} V^2_{\star} ,
      \\[0.2cm]     \label{eq7c}
      & & \hspace{-1.1cm}
P_{sw} = \frac{2 (\gamma-1)}{(\gamma+1)^2} \rho_{sw} V^2_{\star} ,
      \\[0.2cm]     \label{eq7d}
      & & \hspace{-1.1cm}
V_{sw} = \frac{\gamma-1}{\gamma+1} V_{\star} .
\end{eqnarray}
The results of the numerical integration in the case of a $3 
\times 10^5$\Msol , 1~Myr old cluster with $R_{sg} = a$, formed in a 
$10^6$\Msol \, cloud with metallicity $Z = 0.02Z_{\odot}$, are presented in 
Fig. 3. Here the dashed, solid and dotted lines display the distribution of 
the hydrodynamical variables in the subsonic shocked wind zone around an
individual massive star when it is located in the center of 
clusters with different core radii (2~pc, 3~pc and 4~pc, respectively).
Vertical lines indicate the half distance between neighboring massive stars
in each of these cases. The gas, whose speed behind the reverse shock is 
equal to one quarter of the stellar wind terminal speed (in this case 
$V_{\star} = 2370$~km s$^{-1}$), rapidly slows down (see panel a)
because of the inverse thermal pressure gradient in the shocked wind zone. 
When the velocity of the gas becomes negligible and its ram pressure vanishes, 
the thermal pressure reaches a balance with the ambient gas pressure and 
remains homogeneous throughout the shocked wind zone (see panel b). 
In the case of the less compact cloud (dotted lines in Fig. 3), the impact of
radiative cooling is negligible. In this case 
the density and the temperature in the shocked wind zone do not change 
significantly and the shocked wind remains hot as it merges with a neighboring
shocked wind bubble (see dotted lines on panels c and d). Neighboring hot 
bubbles then fill in the whole star-forming region and form a star cluster 
wind able to expel all gas, which includes the residual and that reinserted by
massive stars, from the star cluster volume. Such clusters are not able 
to form a second stellar generation unless they further accrete a sufficient 
amount of matter. One can estimate the characteristic time $\tau_w$, which is 
required for hot bubbles to merge in such clusters, by integrating equation 
(\ref{A6}) from Appendix A. This time depends on the location of the 
neighboring stars in the cluster. In the case of a $3 \times 10^5$\Msol \, 
cluster with a core radius $a = 4pc$ (dotted lines in Fig. 3), $\tau_w \approx
8 \times 10^5$~yr when stars are located near the star cluster center and 
$\tau_w \approx 6 \times 10^5$~yr when they are located at the edge of the 
mass segregation zone. Thus, in this case it takes less than 1~Myr for massive
stars to fill in the star forming region with a hot gas and form a cluster 
wind similar to that suggested by \citet{Chevalier1985}.

However, in more compact or more massive clusters radiative cooling becomes
a dominant factor. In such clusters the gas velocity in the shocked wind zone 
behind the reverse shock also drops rapidly and thermal pressure reaches a 
balance with the pressure in the ambient intra-cloud medium (see solid and 
dashed lines in panels a and b). However, stellar winds heated at the reverse 
shocks behave adiabatically only in a narrow zone behind the shock. At larger
distances radiative losses become important. The shocked gas temperature 
begins to deviate from the post-reverse shock value and drops dramatically 
when the gas in the shocked wind zone becomes thermally unstable (see dashed 
and solid lines on panel d). The shocked wind density increases then orders of 
magnitude (see panel c) to sustain pressure balance, and dense molecular 
clumps are formed. This results in a dynamical, multi-phase intra-cluster 
medium where hot gas is confined to small compact bubbles around individual 
stars and co-exists with partially photo-ionized residual gas and dense clumps
resulting from the thermally unstable reinserted matter as also occurs
in the inner zones of thermally unstable star cluster winds 
\citep[e.g.][]{Wunsch2009}.

Note, that in clusters which are not sufficiently compact or massive, the 
shocked winds may cool before merging if they are in the central zone of the 
cluster, but merge at the edge of the mass segregation zone, as is the 
case in the $3 \times 10^5$\Msol \, clusters with core radii 2~pc and 3~pc 
where the shocked winds cool before merging in the center (see Fig. 3), but 
not at the edge of the mass segregation zone. Hot shocked bubbles never merge 
if the hot shocked gas cools catastrophically at half the distance or smaller 
between neighboring stars located at the edge of the mass segregation zone. 
The negative stellar feedback in such clusters is dramatically reduced. The
hot gas occupies only a fraction of the star-forming volume and is not able to
escape from the potential well of the cluster and form a cluster wind. In 
such clusters the leftover gas may mix with the cool reinserted matter and 
form another generation of stars with a different chemical composition.

The critical star-forming cloud size is determined then by the condition that 
the shocked wind gas at the edge of the mass segregation zone cools down 
exactly at the half distance between neighboring sources. The critical 
half-mass radius $R_{crit}$ of the cluster depends on the mass of the 
star-forming cloud, the star formation efficiency, the natal gas metallicity 
and on the concentration of massive stars towards the star cluster center (on 
the value of the mass segregation radius $R_{sg}$). It also depends on the 
evolutionary time $t$ as the stellar mechanical luminosity, the mass loss 
rate and the number of massive stars change with time (see Fig. 2). One 
can obtain the value of $R_{crit}$
by iterating the star-forming cloud core radius in the initial conditions 
(\ref{eq7a}) - (\ref{eq7d}) and integrating the set of the hydrodynamic 
equations (\ref{eq6a}) - (\ref{eq6c}) numerically until the shocked gas at the
edge of the mass segregation zone cools exactly at the half distance between 
neighboring sources. The critical half-mass radii calculated at different 
times for proto-stellar clouds with masses $10^5$\Msol, $10^6$\Msol \, and 
$10^7$\Msol \, (dotted, solid and dashed lines, respectively) assuming a star 
formation efficiency $\epsilon = 0.3$ and metallicity $Z = 0.02Z_{\odot}$ are 
shown in Fig. 4.
\begin{figure}
\includegraphics[width=\columnwidth]{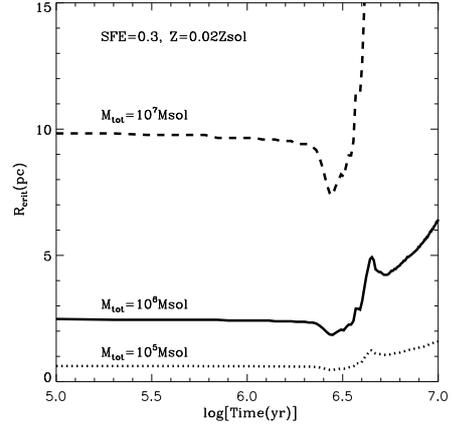}
\caption{The star-forming cloud critical half-mass radii. 
         The calculations were provided for star-forming clouds with 
         $10^5$\Msol \, $10^6$\Msol \, and $10^7$\Msol proto-stellar
         clouds with star formation efficiency $\epsilon = 0.3$, 
         metallicity $Z = 0.02Z_{\odot}$ and the mass segregation radius
         equal to the star-forming cloud core radius, $R_{sg} = a$.} 
 \label{f4}
\end{figure}

$R_{crit}$ reaches the minimum value $R_{min}$ just before the onset of SN 
explosions, when the stellar winds power reaches the maximum value (see 
Fig. 2). After the beginning of SN explosions the critical half-mass radius 
increases rapidly because the power of stellar winds drops and the mean 
separation between massive stars grows. Clusters whose radii are smaller than 
$R_{min}$ retain the residual and the reinserted matter whereas those with 
$R_{hm} > R_{min}$ form hot star cluster winds.

\section{Catastrophic cooling versus the residual gas expulsion}
\label{sec4}

\subsection{The critical size of the star-forming cloud and the 
           critical gas central density}

One can now use the procedure described in the previous section to 
calculate the critical half-mass radius at successive evolutionary times $t$ 
and select the minimum half-mass radius from those calculated at different 
star cluster ages. This determines the critical half-mass radius $R_{hm,crit} 
= R_{min}$ of the star-forming cloud and its emerging cluster. Clusters whose 
radii are smaller than $R_{hm,crit}$ are not able to form a cluster wind and 
expel gas from the star-forming cloud whereas those whose radii are larger 
than $R_{hm,crit}$ form hot winds which clean up the cluster from the residual
and the reinserted matter.

Critical lines which separate clusters that retain gas and those which 
form cluster winds are shown in Fig. 5. Here the results of the calculations 
for clouds with solar $Z = Z_{\odot}$ and $Z = 0.02Z_{\odot}$ abundances are 
shown in the left and right panels, respectively. The mass segregation radius 
selected for these calculations is equal to the star cluster core radius: 
$R_{sg} = a$.
\begin{figure*}
\vspace{16.5cm}
\includegraphics{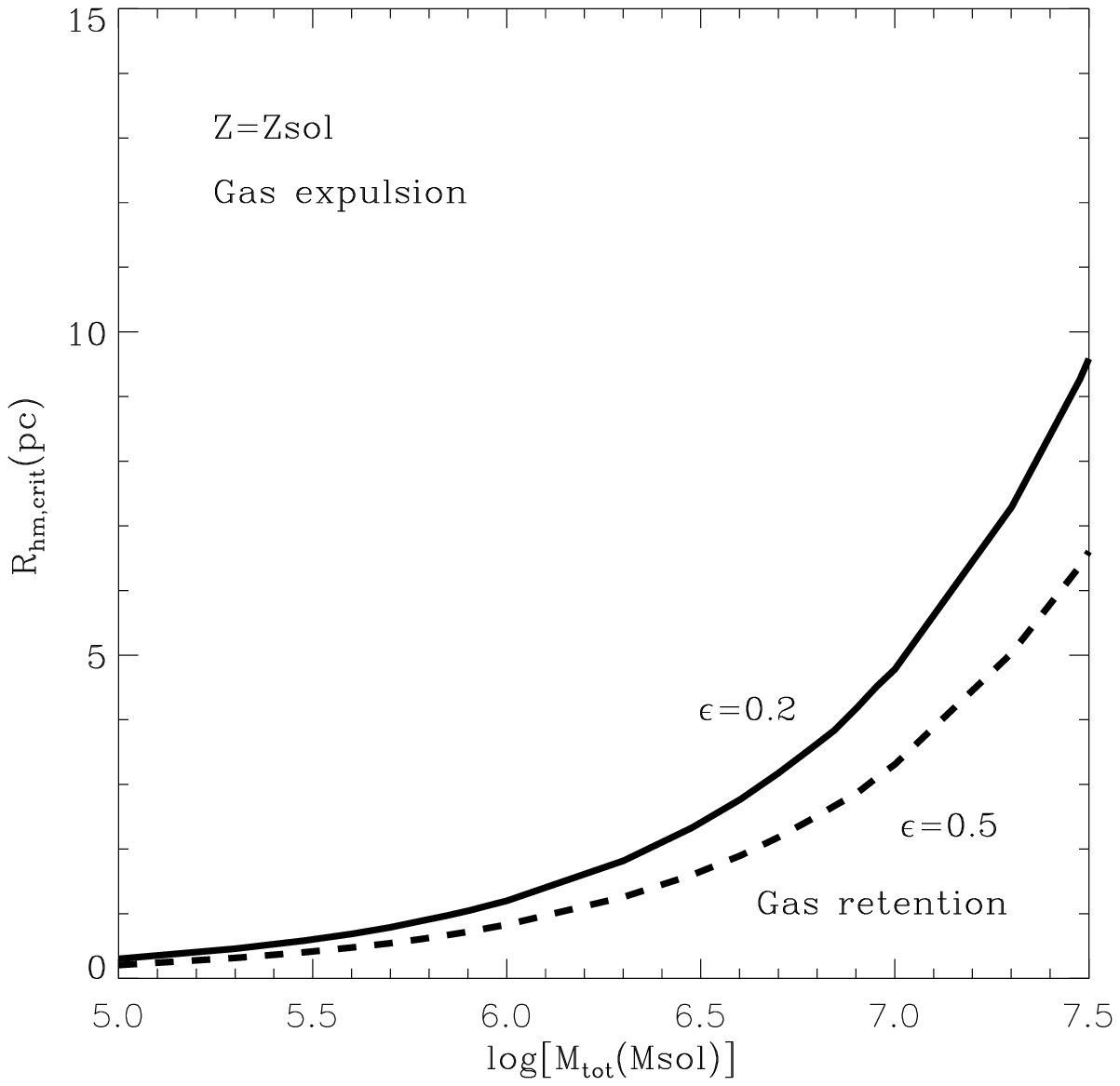}
\includegraphics{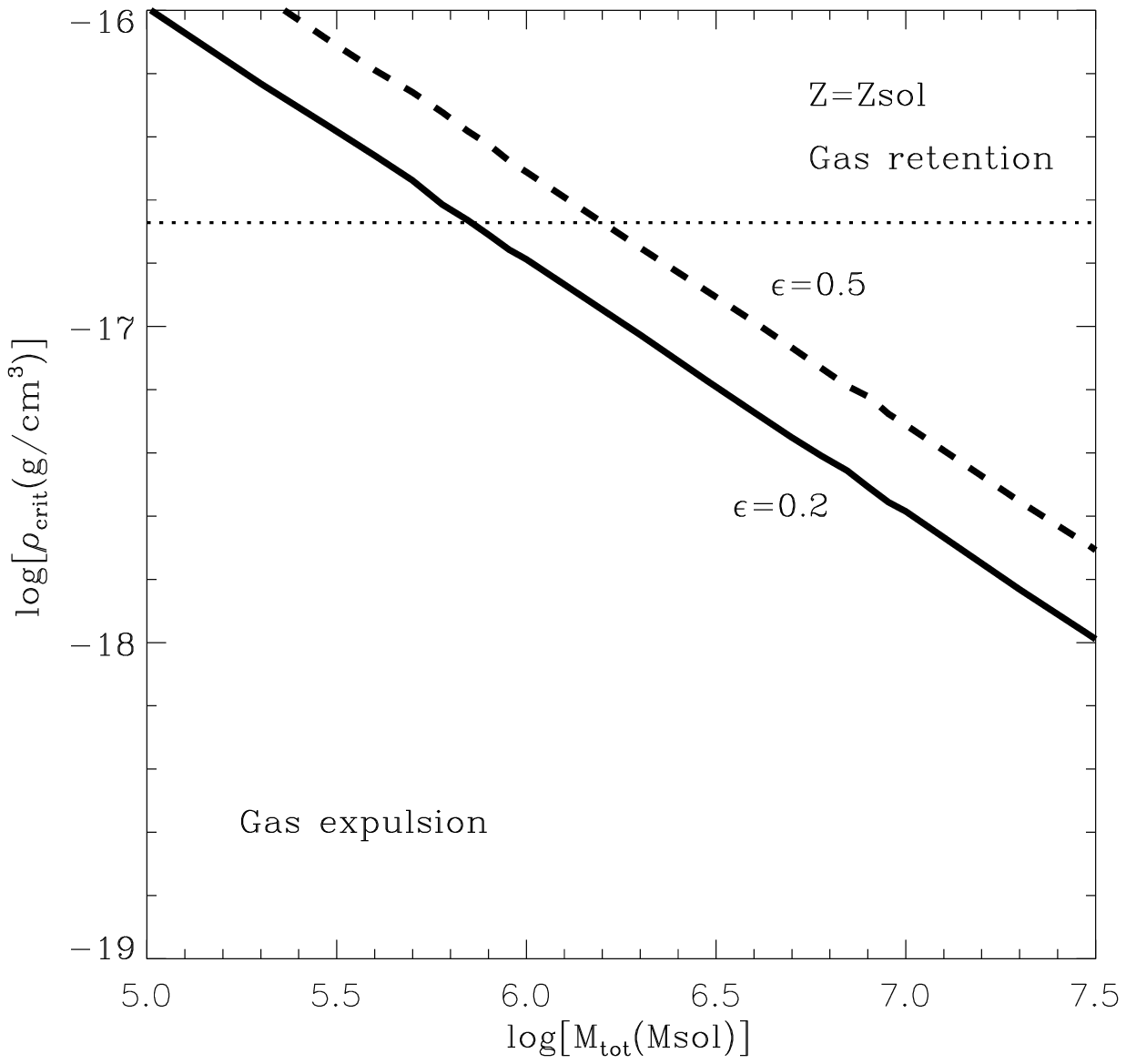}
\includegraphics{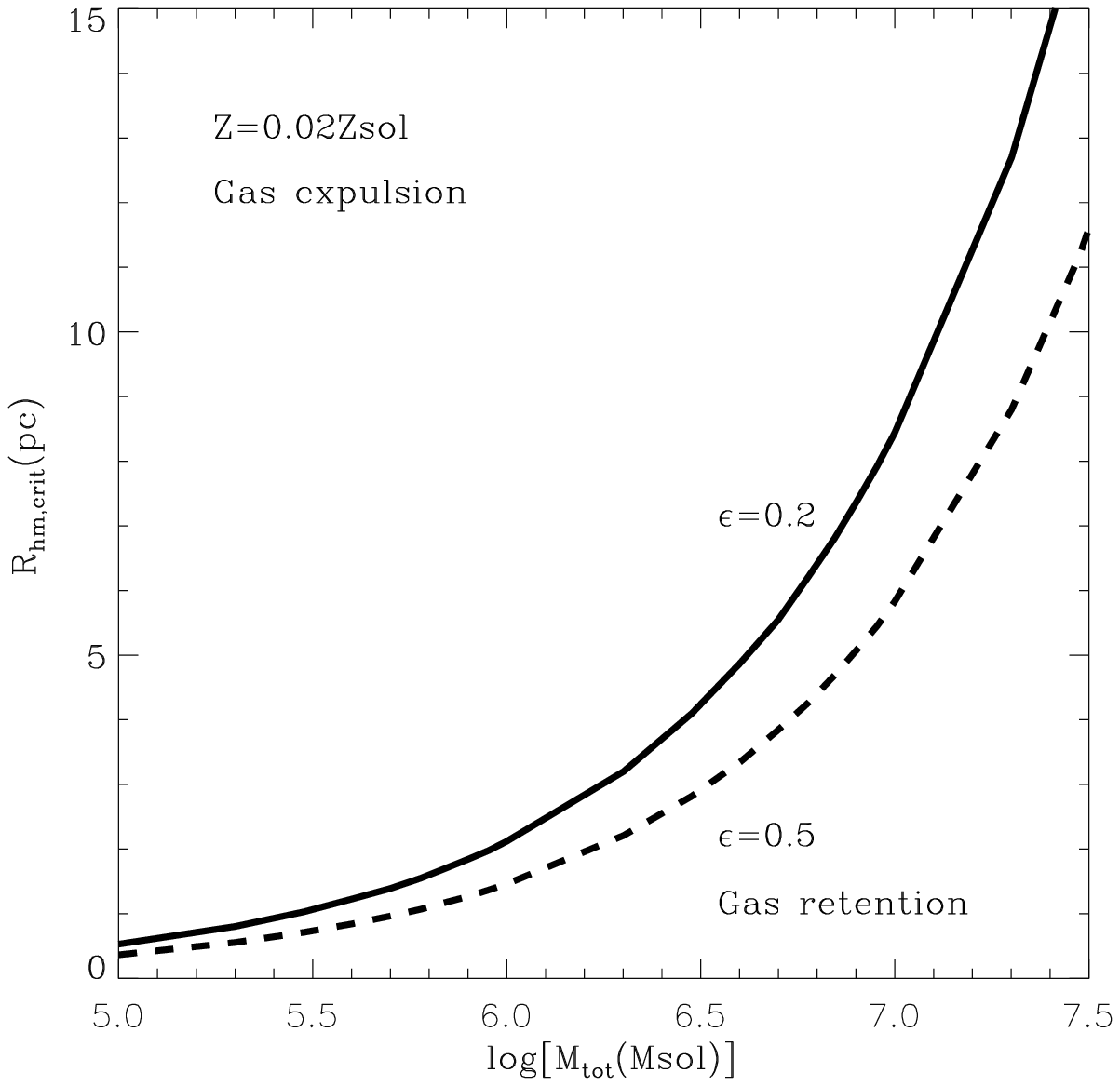}
\includegraphics{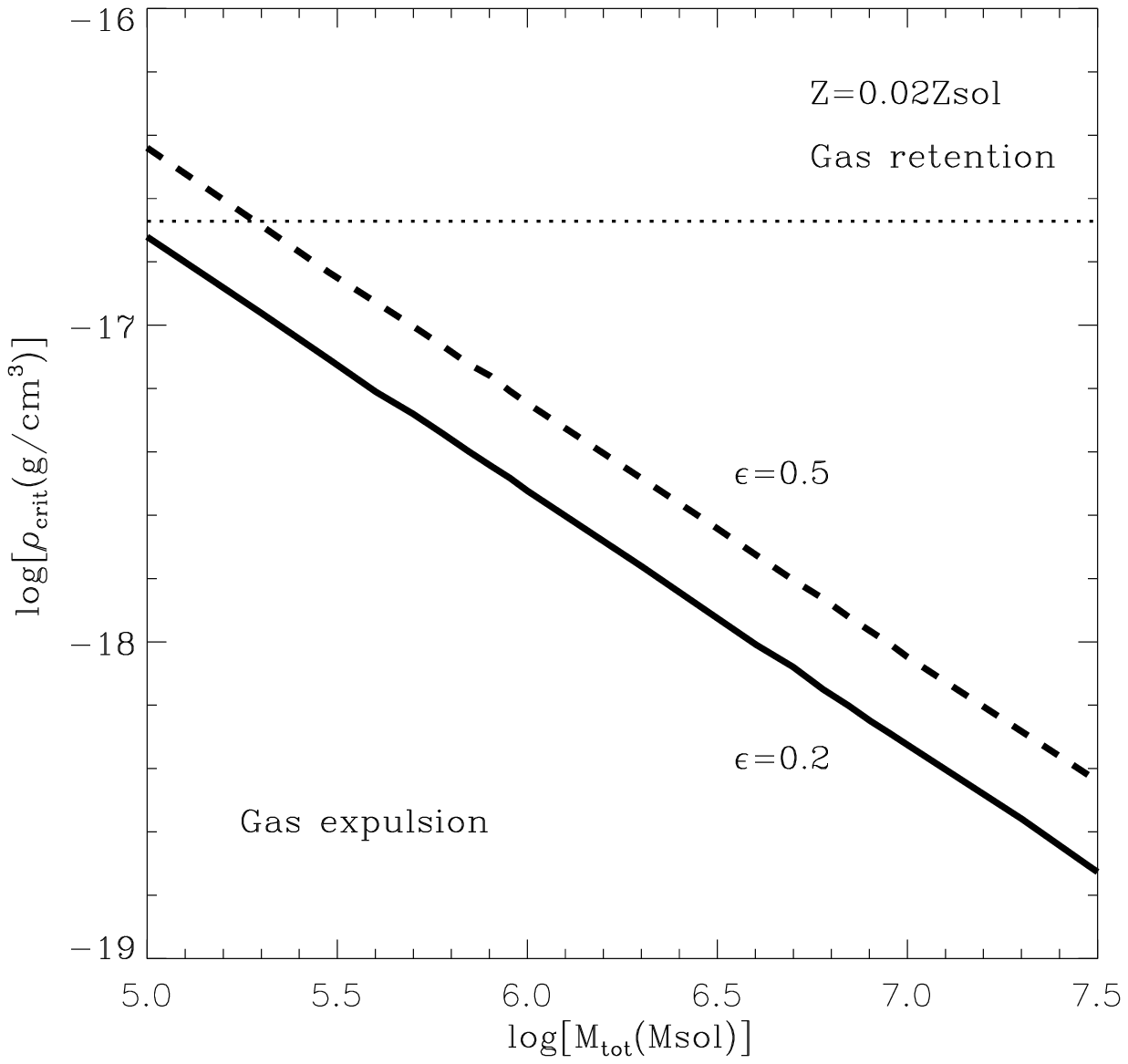}
\caption{Star-forming clouds critical radii and gas central densities.
         The left-hand and right-hand panels present the results of the
         calculations for star-forming clouds with solar and $Z=0.02Z_{\odot}$
         abundances. The upper and bottom panels display the critical 
         half-mass radii and gas central densities, respectively. The solid 
         and dashed lines correspond to star-forming clouds with different
         star formation efficiencies, $\epsilon=0.2$ and $\epsilon=0.5$,
         respectively. The dotted lines on the bottom panels mark the
         largest stellar mass density detected in globular clusters.
         Clusters located below the critical lines on the top panels or
         above the critical lines on the bottom panels retain the leftover
         gas after the 1G formation whereas clusters located above the
         threshold lines on the top panels and below the threshold lines
         on the bottom panels rapidly form star cluster winds.}    
\label{f5}
\end{figure*}
The solid and dashed lines show the critical lines calculated for 
star formation efficiencies $\epsilon = 0.2$ and $\epsilon = 0.5$,
respectively, which are often suggested as lower and upper limits for the 
star formation efficiency in proto-globular clusters \citet{Ashman2001,
Kroupa2001B,Johnson2015}. The upper panels show the critical half-mass radii, 
whereas the bottom ones the critical gas central densities. 
One can notice, that star-forming clouds must be sufficiently massive in order
to be located in the gas retention parameter space and have central gas 
densities which do not exceed the maximum stellar mass and molecular gas 
density ($\sim 10^7$ cm$^{-3}$, the horizontal dotted lines in Fig. 5) so 
far detected in globular clusters and Galactic molecular clouds 
\citep [see][]{Renzini2013,Rathborne2015}. This may explain observational 
constraints on the low mass limit for stellar 
clusters with multiple stellar populations \citep[see][end references therein]
{Carretta2010,Salinas2015,Mucciarelli2016,Massari2017}. It is also remarkable 
that the critical half-mass radii required for the gas retention (see upper 
panels in Fig. 5) are close to those of globular clusters 
\citep{Ashman2001,Johnson2015}. 

A comparison of the left and right panels in Fig. 5 also shows that clusters
formed in present day galaxies must be more massive and compact than young 
proto-globular clusters formed from pristine matter with a low metal 
content in order to retain gas within the star-forming volume. Therefore the 
window of opportunity to form a 2G of stars is larger in the case when star 
formation occurs in low metallicity clouds. 

Fig. 5 also illustrates the significant role of the star formation efficiency 
for the success of gas retention or expulsion from the natal star-forming 
cloud. Indeed, the star-forming cloud location on Fig. 5 depends not only on 
the star cluster mass and compactness, but also on the star formation 
efficiency which determines the amount of gas left over from the 1G formation.

\subsection{Effects of mass segregation}

In order to examine how the concentration of massive stars towards the star 
cluster center affects the critical lines, we calculated the critical radii 
and gas central densities for star-forming clouds with 
different mass segregation radii: $R_{sg} = a$, $R_{sg} = 2 a$ and 
$R_{sg} = a/2$. The results of the calculations are presented in Fig. 6. 
Critical radii and densities do not change significantly in the case when
the mass segregation radius is smaller than the star cluster core radius, 
$R_{sg} < a$ (compare the solid and dotted lines in Fig. 6). However, in the
case when $R_{sg} > a$ (dashed lines in Fig. 6), the difference is notable. 
The impact of the mass segregation zone size on the threshold lines in these 
two cases is different because the ambient gas density and the gas pressure in 
the central zone of the cloud ($r < a$) do not change so rapidly as they do 
in the outer zone (see Fig. 1). In clusters with $R_{sg} < a$ reverse
shocks around massive stars at the edge of the mass segregation zone have 
smaller radii than in the case when $R_{sg} = a$ as the intra-cloud gas 
pressure grows towards the center. This results in larger densities and faster
gas cooling in the shocked wind zones. Therefore in the case when 
$R_{sg} < a$ catastrophic gas cooling occurs in clusters with larger half-mass
radii than in the case when $R_{sg} = a$. The difference between the critical 
lines in this case is negligible as the pressure gradient in the central 
zone of the star-forming cloud is small. In clusters with $R_{sg} > a$ the 
ambient gas pressure at the edge of the mass segregation zone is smaller than 
in clusters with $R_{sg} = a$. Therefore reverse shocks are located further 
away from their stars and star-forming clouds must be more compact in order 
that shocked winds cool before merging with their neighbors. In this case the 
difference between the critical lines is significant as the intra-cloud 
pressure drops rapidly outside of the central zone and the star-forming cloud 
must be much more compact than in the case when $R_{sg} = a$ in order the 
shocked winds at the edge of the mass segregation zone cool before merging 
with their neighbors.
\begin{figure}
\includegraphics[width=\columnwidth]{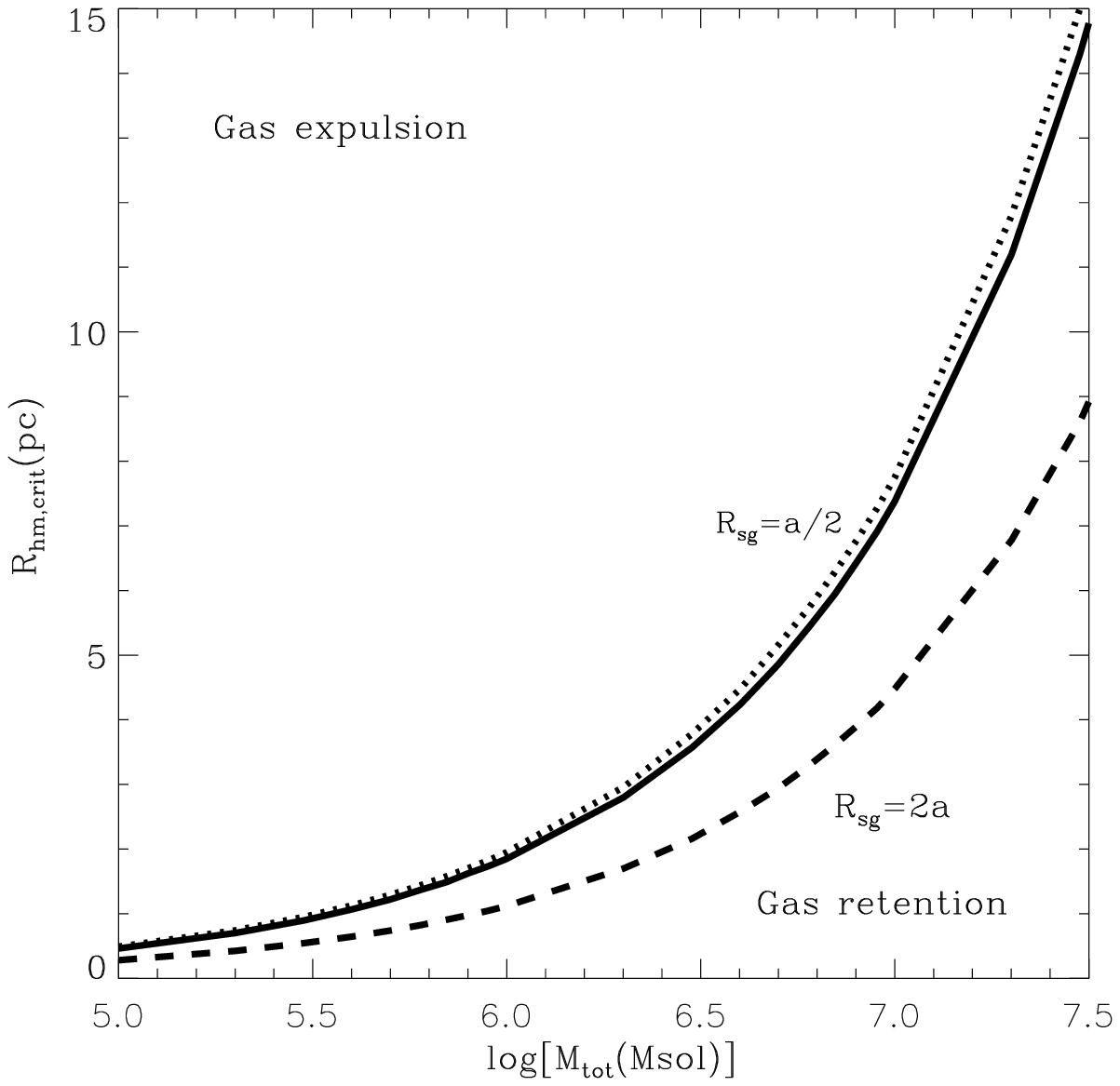}
\includegraphics[width=\columnwidth]{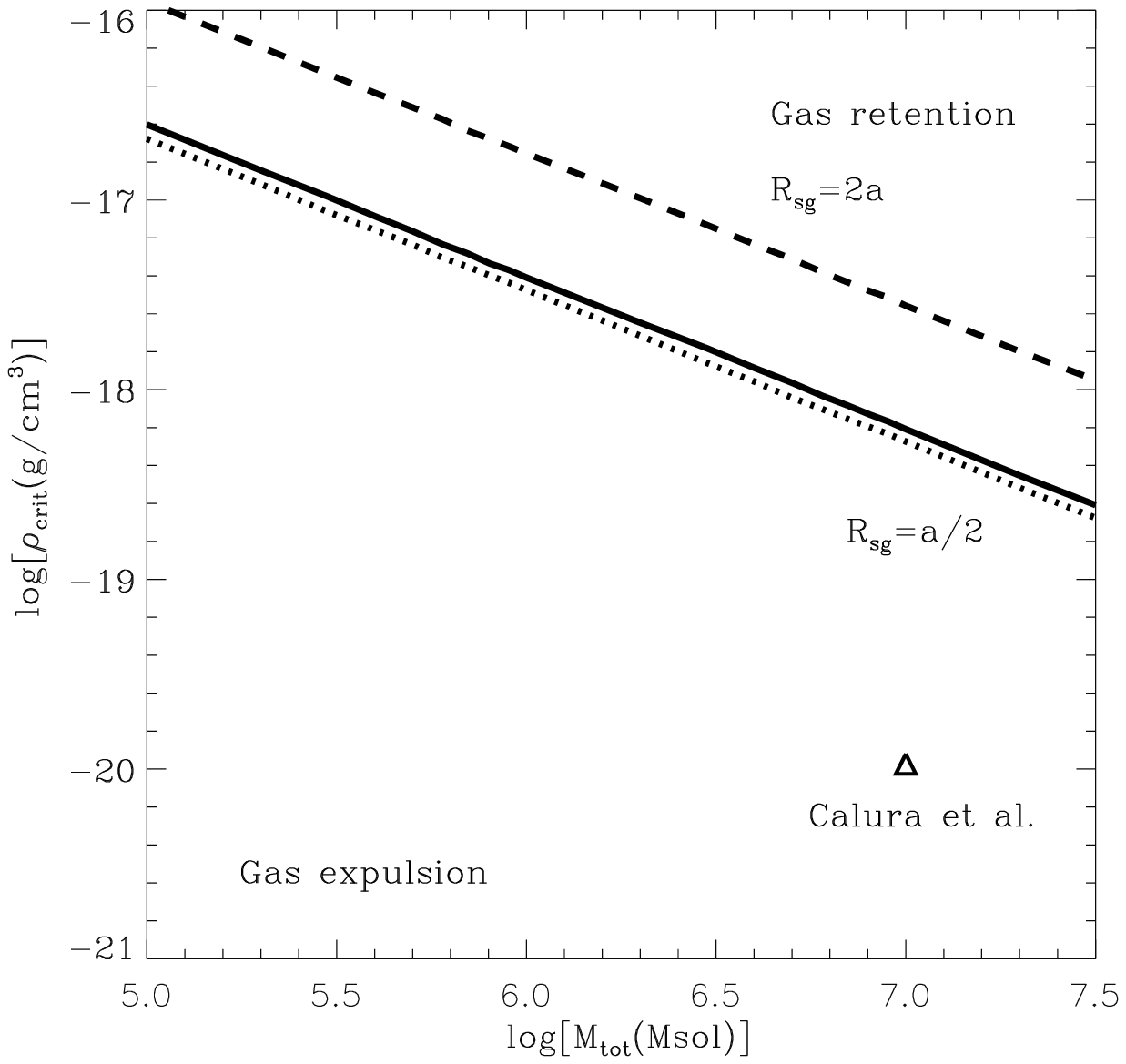}
\caption{Critical lines for clusters with different mass segregation
         radii. The solid, dotted and dashed lines displays the critical 
         lines derived for clusters with the same metallicity
         ($Z = 0.02Z_{\odot}$) and the star formation efficiency 
         ($\epsilon = 0.3$) but different mass segregation radii: 
         $R_{sg} = a$, $R_{sg} = a/2$ and $R_{sg} = 2a$, respectively.}
\label{f6}
\end{figure}
The triangle symbol in Fig. 6 marks the initial gas central density used
by \citet[][]{Calura2015} in their 3D numerical simulations. One can note 
that clouds with such star formation efficiency ($\epsilon = 0.3$) and such
gas central densities ($\sim 5 \times 10^3$cm$^{-3}$) are located  in the gas 
expulsion parameter space, far away from the critical lines. Thus the 
conclusion made by \citet[][]{Calura2015} who claimed that stellar winds 
remove the residual gas from such star-forming cloud is in agreement 
with our findings.

\subsection{Star cluster compactness and critical star formation 
            efficiency}

Top panels in Fig. 5 and Fig. 6 also show that only sufficiently compact 
clusters are able to retain the leftover and the reinserted matter. The 
crucial role of the cluster compactness was also stressed by 
\citet{Krause2016} who introduced the star cluster compactness index 
$C_5 = (M_{SC}/10^5\Msol)/(R_{hm}/1pc)$ and concluded that this index together
with star formation efficiency completely determine the success of the gas 
expulsion from the star-forming cloud  (see their Fig. 4). 

In our model the compactness of star-forming clouds is also important. 
However the critical efficiency depends not only on the compactness index, but 
also on the star cluster mass.  The critical star formation efficiencies 
calculated for clusters with $Z_{\odot}$ and $0.02Z_{\odot}$ abundances 
are shown on the top and bottom panels in Fig. 7. Here the solid, dotted and 
dashed lines display critical star formation efficiencies for $10^5$\Msol , 
$10^6$\Msol \, and  $10^7$\Msol \, clusters, respectively. The star formation 
efficiency must be smaller than the critical one if 1G stars are to avoid 
forming global wind and the star cluster retains the leftover gas and the
reinserted matter. One can note that less massive clusters with a wide 
spread of compactness index and very high star formation efficiencies may
retain the residual gas. In such clusters the stellar feedback is ineffective 
and they essentially evolve in the gas exhaustion regime discussed in 
\citep[][]{Ginsburg2016,Longmore2014}. 
\begin{figure}
\includegraphics[width=\columnwidth]{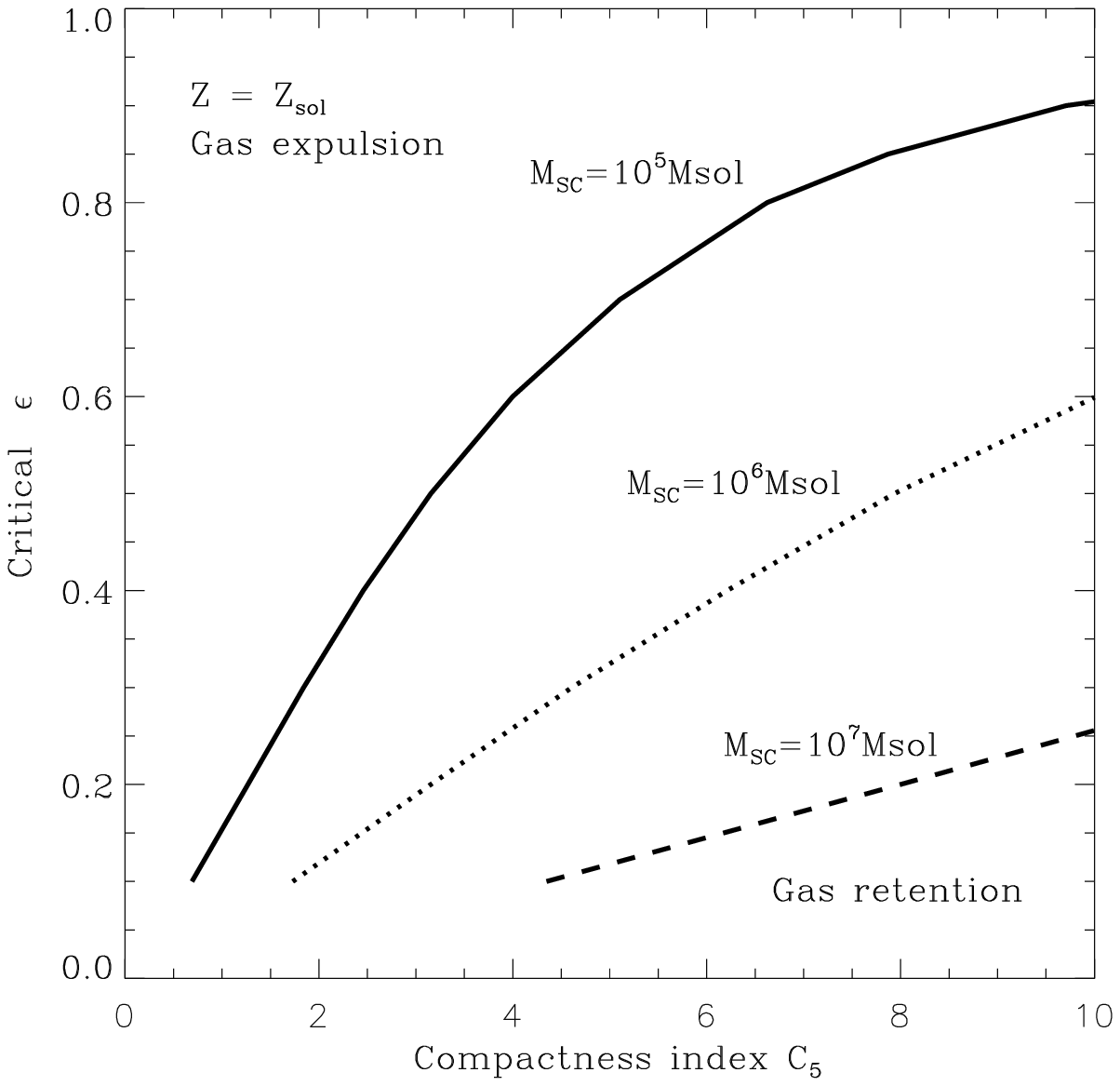}
\includegraphics[width=\columnwidth]{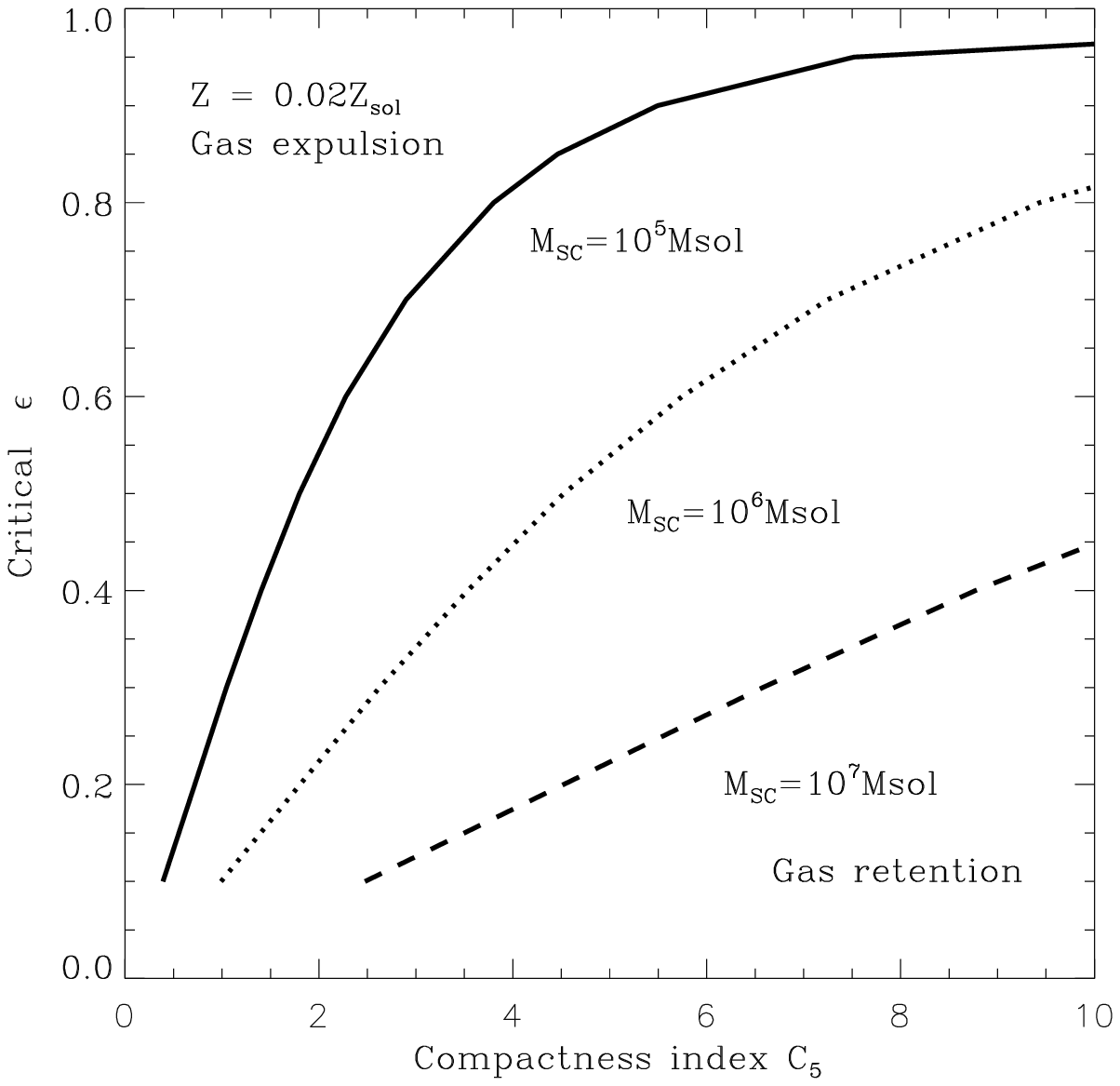}
\caption{Critical 1G star formation efficiency required for gas retention in
young stellar cluster with a given compactness index $C_5$. The top and bottom
panels present critical star formation efficiencies calculated for clusters
with solar and $0.02Z_{sol}$ abundances, respectively. The solid, dotted and 
dashed lines show critical star formation efficiencies calculated for 
$10^5$\Msol, $10^6$\Msol \, and $10^7$\Msol \, clusters, respectively. 
Clusters with a given compactness index $C_5$ retain gas if the star formation 
efficiency in a star-forming cloud is smaller than the critical value. The 
global wind is formed, the leftover and the reinserted gas are expelled from 
the cluster if the star formation efficiency is larger than the critical one. 
The calculations were provided for low metallicity ($Z = 0.02Z_{\odot}$) 
clusters with $R_{sg} = a$.}
\label{f7}
\end{figure}

\section{Conclusions}

Strong negative feedback from massive stars is usually suggested to be 
responsible for expelling the residual gas and terminating star formation
in young stellar clusters. Here we show that negative stellar feedback, which 
leads to the star-forming cloud destruction, is strongly suppressed in massive
and compact star-forming regions. The critical half-mass radii and gas central
densities which separate clusters evolving in the negative and positive 
feedback regimes are obtained. They depend on the star formation efficiency 
and metallicity of the star-forming cloud and on the primordial mass 
segregation in the assembling cluster.

In clusters with a negative stellar feedback individual shocked winds merge
to form star cluster winds, clear star-forming regions and terminate star
formation in a short time-scale. The further star formation in such clusters 
is possible only after all massive stars from the 1G terminate their life 
cycle and requires the ambient interstellar medium to be accreted at later 
stages of their evolution. 

In clusters with a positive star formation feedback hot shocked winds around
individual massive stars cool before merging with their neighbors. Such 
clusters do not form hot star cluster winds which expel gas from their
parental clouds. They are invisible in the optical or UV emission being deeply 
embedded into their dense molecular, partially ionized clouds and a system of
dense clumps containing thermally unstable processed matter. 
Such deeply embedded clusters should be detected due to their powerful 
radio-continuum and IR radiation coincident with the molecular gas emission as
it is the case in the NGC5253 supernebula \citep{Turner2000,Gorjian2001,
Turner2003,Beck2012,Beck2015,Turner2015}. It is likely that in such clusters 
further generations of stars with different chemical abundances to be formed.

The window of opportunity for clusters to retain gas and form a second stellar 
population is larger when they develop in clouds with low abundances. In this
case less compact clusters formed in less dense proto-cluster clouds may
evolve in the positive star formation regime.

It is remarkable that the model-predicted half-mass radii which separate 
clusters evolving in the negative and positive star-formation regimes 
are similar to those of globular clusters. Small sizes and high central 
densities required for the gas retention and secondary star formation 
suggest that clusters with a positive star formation feedback are formed in a 
high pressure environment. Such conditions together with low abundances 
favor early assembling galaxies to be the nests of present-day globular 
clusters with multiple stellar populations.

\section*{Acknowledgments}

The authors thank S. Cassisi for his comments regarding the low mass limit
for globular clusters with multiple stellar populations and their anonymous 
referee for a detailed report full of suggestions which have largely improved 
the presentation of their results. 

\bibliographystyle{mnras}
\bibliography{GC}

\appendix

\section{Subsonic growth of a hot shocked zone}

Hot, shocked wind zones continue to grow even after the gas pressure drops to 
that in the turbulent ambient medium and the wind-driven leading shock 
vanishes, as their central stars continue to deposit mass and energy via 
their stellar winds. The shocked wind zone grows then in a sub-sonic regime 
remaining in an approximate pressure equilibrium with the ambient intra-cloud
medium. As the velocity of the shocked gas is much smaller than its sound
speed (see Fig. 3), one can consider only the shocked wind thermal energy.
The subsonic growth of the shocked wind zone is determined then by the
following set of equations: 
\begin{eqnarray}
      \label{A1}
          & & \hspace{-1.1cm} 
\der{E_{th}}{t} = L_{\star} - 4\pi P u R^2 ,
      \label{A2}
          \\[0.2cm]      
          & & \hspace{-1.1cm} 
u = \der{R}{t},
       \label{A3}
          \\[0.2cm]      
          & & \hspace{-1.1cm}
E_{th} = \frac{4\pi P}{3(\gamma-1)} R^3 ,
       \label{A4}
          \\[0.2cm]      
          & & \hspace{-1.1cm}
P = P_g ,
\end{eqnarray}
where $E_{th}$, $R$ and $u$ are the thermal energy, radius and growth velocity
of the shocked wind zone, respectively. $P_g$ is the intra-cloud gas pressure,
which is determined by equation (\ref{eq1d}) and $L_{\star}(t)$ is a stellar
wind mechanical power determined in section 2.2. Because mean 
separations between neighboring stars and radii of the shocked wind zones are 
small, one can adopt that the confining pressure $P_g$ does not change as the 
shocked wind zone grows up. The thermal energy $E_{th}$ and the shocked zone 
growth velocity $u$ could be eliminated from the set of equations 
(\ref{A1})-(\ref{A4}) yielding: 
\begin{equation}
      \label{A6}
\der{R}{t} = \frac{(\gamma-1) L_{\star}(t)}
                 {4 \pi \gamma P_t} R^{-2} . 
\end{equation}
The initial conditions to solve this equations are: $R(t_0) = R_{stall}$,
$t_0 = \tau_{stall}$, where 
the wind-driven shell stalling radius $R_{stall}$ and stalling time 
$\tau_{stall}$ are:
\begin{eqnarray}
      \nonumber
      & & \hspace{-0.9cm} 
R_{stall} = \left(\frac{14 a^2}{25 G}\right)^{3/4} 
\left(\frac{3 a^3}{M^5_{tot}}\right)^{1/4}
              \left[\frac{375(\gamma-1) L_{\star}}
               {7 (9\gamma-4) (1 - \epsilon)}\right]^{1/2}
      \\[0.2cm]     \label{A8}
      & & \hspace{0.5cm}
\times \left(1 + \frac{r^2}{a^2}\right)^{13/8} ,
      \\[0.2cm]  \nonumber
      & & \hspace{-0.9cm}
\tau_{stall} = \frac{R_{stall}}{a} 
     \left[\frac{7(9\gamma-4)(1-\epsilon) M_{tot} R_{stall}^2}
     {125(\gamma-1) L_{\star}}\right]^{1/3} 
      \\[0.2cm]     \label{A9}
      & & \hspace{0.5cm}
\times     \left(1 + \frac{r^2}{a^2}\right)^{-5/6} .
\end{eqnarray}

One can solve equation (\ref{A6}) numerically by making use the wind-driven
shell stalling time and stalling radius as the initial conditions for the
numerical integration. Note, that in the case of a constant mechanical 
luminosity, equation (\ref{A6}) has an analytic solution:
\begin{equation}
     \label{A10}
R(t) = R_{stall} \left(1 + \frac{t}{\tau_0}\right)^{1/3} , 
\end{equation}
where $t$ is the time since the wind-driven shell stalls and
\begin{equation}
      \label{A11}
\tau_0 = \frac{4 \pi \gamma P_g R^3_{stall}}{3 (\gamma-1) L_{\star}} .
\end{equation}
  
\bsp	
\label{lastpage}
\end{document}